\documentclass[fleqn,10pt]{wlscirep}
\usepackage[utf8]{inputenc}
\usepackage[T1]{fontenc}
\usepackage{bm}
\usepackage{xr}
\usepackage{cleveref}
\usepackage{mathtools}
\usepackage[mathscr]{euscript}
\usepackage{float}
\usepackage{url}
\usepackage{csquotes}
\usepackage[backend=biber,style=nature,natbib=true, autocite=superscript]{biblatex}  
\renewcommand{\cite}{\autocite}
\usepackage{amsmath}
\usepackage{algorithm}  
\usepackage[noend]{algpseudocode}   
\usepackage[version=4]{mhchem}

\usepackage{bm} 
\newcommand{\bs}[1]{\bm{#1}}

\definecolor{todo}{RGB}{255, 0, 0}
\definecolor{2ndEd}{RGB}{127,0,255}
\definecolor{3rdEd}{RGB}{127,0,255}
\definecolor{moved}{RGB}{138, 154, 91}

\title{Structural Constraint Integration in Generative Model for Discovery of Quantum Material Candidates}
\author[1,2, *]{Ryotaro Okabe}
\author[1,3,4]{Mouyang Cheng}
\author[1,5]{Abhijatmedhi Chotrattanapituk}
\author[1,6,7]{Nguyen Tuan Hung}
\author[5]{Xiang Fu}
\author[8]{Bowen Han}
\author[9]{Yao Wang}
\author[10]{Weiwei Xie}
\author[11]{Robert J. Cava}
\author[5]{Tommi S. Jaakkola}
\author[8,**]{Yongqiang Cheng}
\author[1,6,***]{Mingda Li}
\affil[1]{Quantum Measurement Group, Massachusetts Institute of Technology, Cambridge, MA, USA}
\affil[2]{Department of Chemistry, Massachusetts Institute of Technology, Cambridge, MA, USA}
\affil[3]{Department of Materials Science and Engineering, Massachusetts Institute of Technology, Cambridge, MA, USA}
\affil[4]{Center for Computational Science $\&$ Engineering, Massachusetts Institute of Technology, Cambridge, MA, USA}
\affil[5]{Department of Electrical Engineering and Computer Science, Massachusetts Institute of Technology, Cambridge, MA, USA}
\affil[6]{Department of Nuclear Science and Engineering, Massachusetts Institute of Technology, Cambridge, MA, USA}
\affil[7]{Frontier Research Institute for Interdisciplinary Sciences, Tohoku University, Sendai 980-8578, Japan}
\affil[8]{Chemical Spectroscopy Group, Spectroscopy Section, Neutron Scattering Division Oak Ridge National Laboratory, Oak Ridge, TN, USA}
\affil[9]{Department of Chemistry, Emory University, Atlanta, Georgia, USA}
\affil[10]{Department of Chemistry, Michigan State University, East Lansing, MI, USA}
\affil[11]{Department of Chemistry, Princeton University, Princeton, NJ, USA}

\affil[*]{e-mail: rokabe@mit.edu}
\affil[**]{e-mail: chengy@ornl.gov}
\affil[***]{e-mail: mingda@mit.edu}

\begin{document}

\begin{abstract}
Billions of organic molecules are known, but only a tiny fraction of the functional inorganic materials have been discovered, a particularly relevant problem to the community searching for new quantum materials. Recent advancements in machine-learning-based generative models, particularly diffusion models, show great promise for generating new, stable materials. However, integrating geometric patterns into materials generation remains a challenge. Here, we introduce Structural Constraint Integration in the GENerative model (SCIGEN). Our approach can modify any trained generative diffusion model by strategic masking of the denoised structure with a diffused constrained structure prior to each diffusion step to steer the generation toward constrained outputs. Furthermore, we mathematically prove that SCIGEN effectively performs conditional sampling from the original distribution, which is crucial for generating stable constrained materials. We generate eight million compounds using Archimedean lattices as prototype constraints, with over 10\% surviving a multi-staged stability pre-screening. High-throughput density functional theory (DFT) on 26,000 survived compounds shows that over 50\% passed structural optimization at the DFT level. Since the properties of quantum materials are closely related to geometric patterns, our results indicate that SCIGEN provides a general framework for generating quantum materials candidates.
\end{abstract}
\flushbottom
\maketitle
\thispagestyle{empty}

\section*{Introduction} 
The structure-property relationships are instrumental in understanding quantum and functional materials, and are a fundamental part of any materials science curriculum. Two key structural indicators, symmetry, and geometric pattern, profoundly influence materials properties. For example, materials with inversion symmetry can lead to topological crystalline insulators\cite{Fu2012TCI}, whereas breaking inversion symmetry can result in a variety of phenomena such as Rashba spin-orbit coupling\cite{Bihlmayer2022Rashba}, ferroelectricity\cite{Martin2017FE}, second harmonic generation\cite{Hung2024SHG}, and topological Weyl semimetals\cite{Armitage2018Weyl}. Meanwhile, a material's geometric pattern is closely linked to its electronic states and magnetic orderings. The square lattice serves as a prototype for high-temperature cuprate superconductors\cite{Hashimoto2014}, while triangular, honeycomb, and kagome lattices can host exotic magnetic states like quantum spin liquids \cite{Savar2017QSL,broholm2020quantum}. Additionally, kagome and Lieb lattices can support electronic flat bands \cite{Kang2020CoSn,slot2017experimental} with the technological importance of replacing rare-earth elements\cite{Checkelsky2024}. Also, porous structures like zeolite lattices find applications in catalysis\cite{Speybroeck2015}. However, designing stable materials with desired properties can be nontrivial. For example, only a dozen quantum spin liquid candidates have been identified after a decade of research\cite{Chamorro2021QSLChem}, and even fewer are known for the Lieb lattice.

Machine-learning (ML) based materials generators have led to a paradigm shift in material design. Diffusion models like CDVAE, UniMat, and DiffCSP\cite{xie2021crystal, yang2023scalable,jiao2024crystal}, and graph neural network models like GNoME\cite{merchant2023scaling}, have shown great promise in identifying stable structures to generate millions of materials. However, most ML-based generators create new materials with respect to the distribution of the database, making it challenging to generate materials with specific constraints. Although there have been some developments in the incorporation of crystallographic space groups in the materials generation\cite{jiao2024space, zeni2023mattergen, cao2024space}, the integration of geometric patterns into generation algorithms for functional materials remains challenging. In quantum materials, space group symmetry and geometric patterns can be independent; including the space group symmetry alone will sometimes not allow for proper screening, as is often encountered in monoclinic stacking variants of hexagonal symmetry layers. Moreover, in frustrated magnets, the geometric pattern such as a kagome lattice filled with magnetic atoms plays more important roles than the overall space group in supporting the exotic magnetic structures. Therefore, there is a pressing need to develop an ML-based generator capable of producing new materials constrained by particular geometric patterns.

To answer this need, in this work, we present SCIGEN: Structural Constraint Integration in the GENerative model. SCIGEN is a scheme that can be utilized by any pre-trained generative diffusion model for the incorporation of geometric pattern and symmetry constraints during the generation, without the need for retraining or fine-tuning. Starting from the target constraints, SCIGEN diffuses a random constrained structure over multiple time steps. The constrained structures are used to mask the denoised structure before each diffusion step, creating an inductive bias that directs the generation process toward producing outputs that adhere to the constraints. It turns out that, as we have proven, SCIGEN effectively performs conditional generation with respect to the distribution of the base model. This indicates that the constraint set by SCIGEN would preserve the integrity of the base generative model, including but not limited to the stability of generated materials. To demonstrate, we apply SCIGEN to DiffCSP\cite{jiao2024crystal} for generating materials constrained by Archimedean lattices (ALs) \cite{martinez1973archimedean, eddi2009archimedean}, which are a collection of 2D lattice tiling with square, triangular, honeycomb, kagome, and a few other geometric patterns, and rich harbors for exotic quantum materials. We generate a total of 7.87 million materials belonging to ALs. After a four-staged stability pre-screening, over 790,000 materials survived. Structure relaxation on a subset containing 26,000 materials is computed with high-throughput density functional theory (DFT), showing that 95\% completed the calculation, and more than 53\% can reach the energy minimum within 150 steps of structural optimization. Since SCIGEN requires no extra training apart from the underlying generative model, it also offers a flexible and generically applicable conditioning scheme of materials generation with constraints from both symmetry and geometric patterns. 

\section*{Results}
\subsection*{Structural constraint integration in the generative model}
\begin{figure}[ht!]
\includegraphics[width=0.9\textwidth]{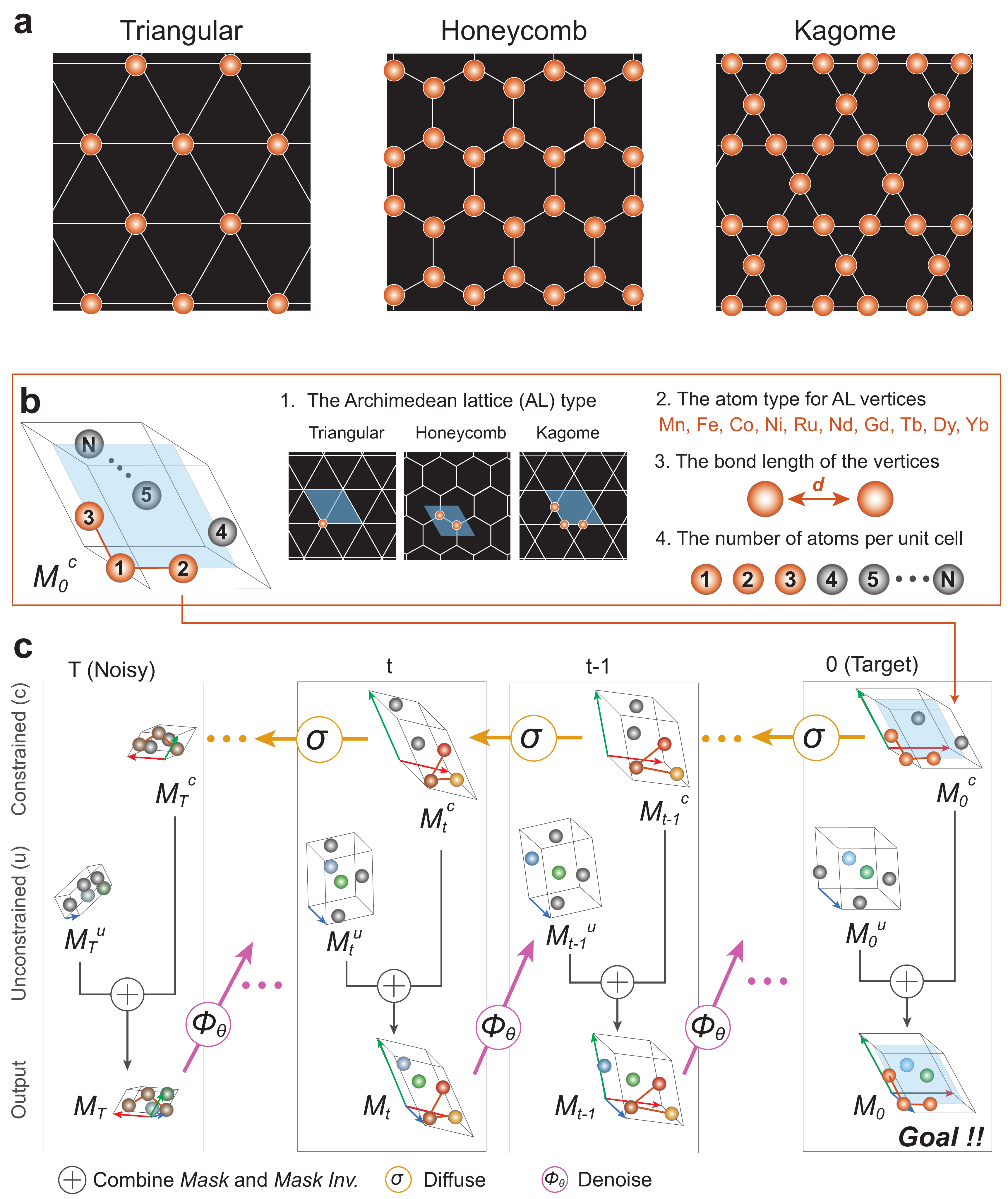}
\centering
\caption{\textbf{Schematic overview of material generation with geometric patterns as constraints.}  
\textbf{a.} Three primary classes of Archimedean lattices with hexagonal unit cells: triangular, honeycomb, and kagome. \textbf{b.} Guideline for structure initialization for diffusion model, with magnetic atoms at Archimedean lattice vertices. Required components include: (1) lattice types, (2) magnetic atom types, (3) nearest-neighbor distances, and (4) total number of atoms per unit cell. \textbf{c.} Methodology of crystal structure generation via diffusion denoising probabilistic model with geometrical pattern as constraints. The initialized structures are iteratively made noisy ($\sigma$), to prepare predefined pathway of the constrained structure $\boldsymbol{M}_{t}^c$, $t \in [1, T]$. For each denoising step $t$, an unconstrained structure $\boldsymbol{M}_{t}^u$ is combined with constrained structure $\boldsymbol{M}_{t}^c$ to get an integrated structure $\boldsymbol{M}_{t}$. $\boldsymbol{M}_{t}$ is passed to the denoising model $\Phi_{\theta}$ and denoised to become the unconstrained structure $\boldsymbol{M}_{t-1}^u$. By repeating this process, we obtain the final crystal structure $\boldsymbol{M}_{0}$, which is guided by the geometrical pattern constraints $\boldsymbol{M}_{0}^c$ but remains realistic with a fair chance to maintain stability. 
}

\label{overview}
\end{figure}

Figure \ref{overview} presents the schematic overview of SCIGEN. The goal of crystal structure generation is to find periodic crystals $\bs{M}$, which can be represented by the three components: the lattice matrix containing three basis vectors $\bs{L}=\left[\bs{l}_{1}, \bs{l}_{2}, \bs{l}_{3}\right] \in \mathbb{R}^{3 \times 3}$, the fractional coordinates $\bs{F}=\left[\bs{f}_{1}, \bs{f}_{2}, \ldots, \bs{f}_{N}\right] \in [0, 1)^{3 \times N}$, and one-hot representations of atom types $\bs{A}=\left[\bs{a}_{1}, \bs{a}_{2}, \ldots, \bs{a}_{N}\right] \in$ $[0, 1]^{h \times N}$. Our methods impose geometric constraints on $\bs{L}$, $\bs{F}$ and $\bs{A}$ respectively in diffusion-based material generation. Figure \ref{overview}a illustrates notable geometries including triangular, honeycomb, and kagome lattices. Following the guideline in Fig. \ref{overview}b, we initiate the constrained structures with an AL composed of magnetic atoms. Figure \ref{overview}c explains the algorithm of the generative process integrating the constrained components. The initialized structure is subjected to a diffusion process by adding noise over $T$-steps denoted as $\boldsymbol{M}_{t}^c$ where $t \in [1..T]$ ($T=1000$ by choice), providing the pre-defined pathway of denoising process for the constrained components. An unconstrained structure is initiated as a completely noisy structure $\boldsymbol{M}_{T}^u$. Both $\boldsymbol{M}_{T}^c$ and $\boldsymbol{M}_{T}^u$ are integrated to form $\boldsymbol{M}_{T}$. This composite structure $\boldsymbol{M}_{T}$ is then denoised to retrieve $\boldsymbol{M}_{T-1}^u$. SCIGEN repeats this process through all steps; it merges $\boldsymbol{M}_{t}^c$ and $\boldsymbol{M}_{t}^u$ to get $\boldsymbol{M}_{t}$, and then predicts $\boldsymbol{M}_{t-1}^u$. This iteration optimizes the final material structure $\boldsymbol{M}_{0}$ by guiding a subset of atoms to form AL planar structures. More details are shown in Supplementary Information 3. Additionally, as proven in Supplementary Information 4, SCIGEN can take the structural constraints and fill the remaining unconstrained components, while preserving the integrity of the unconstrained optimization. Following the generation of a large set of material candidates, we evaluate their stability through a four-staged pre-screening process . The pre-screening involves applying chemical rules such as charge neutrality and the volume of atoms occupying the lattice unit cell, along with auxiliary neural networks that predict stability based on the energy above convex hull ($E_{\text{hull}}$) values. After that, we employ high throughput DFT to relax structures and identify potentially stable candidates.

\subsection*{Materials generation with Archimedean lattice constraints}

\begin{figure}[ht!]
\includegraphics[width=0.9\textwidth]{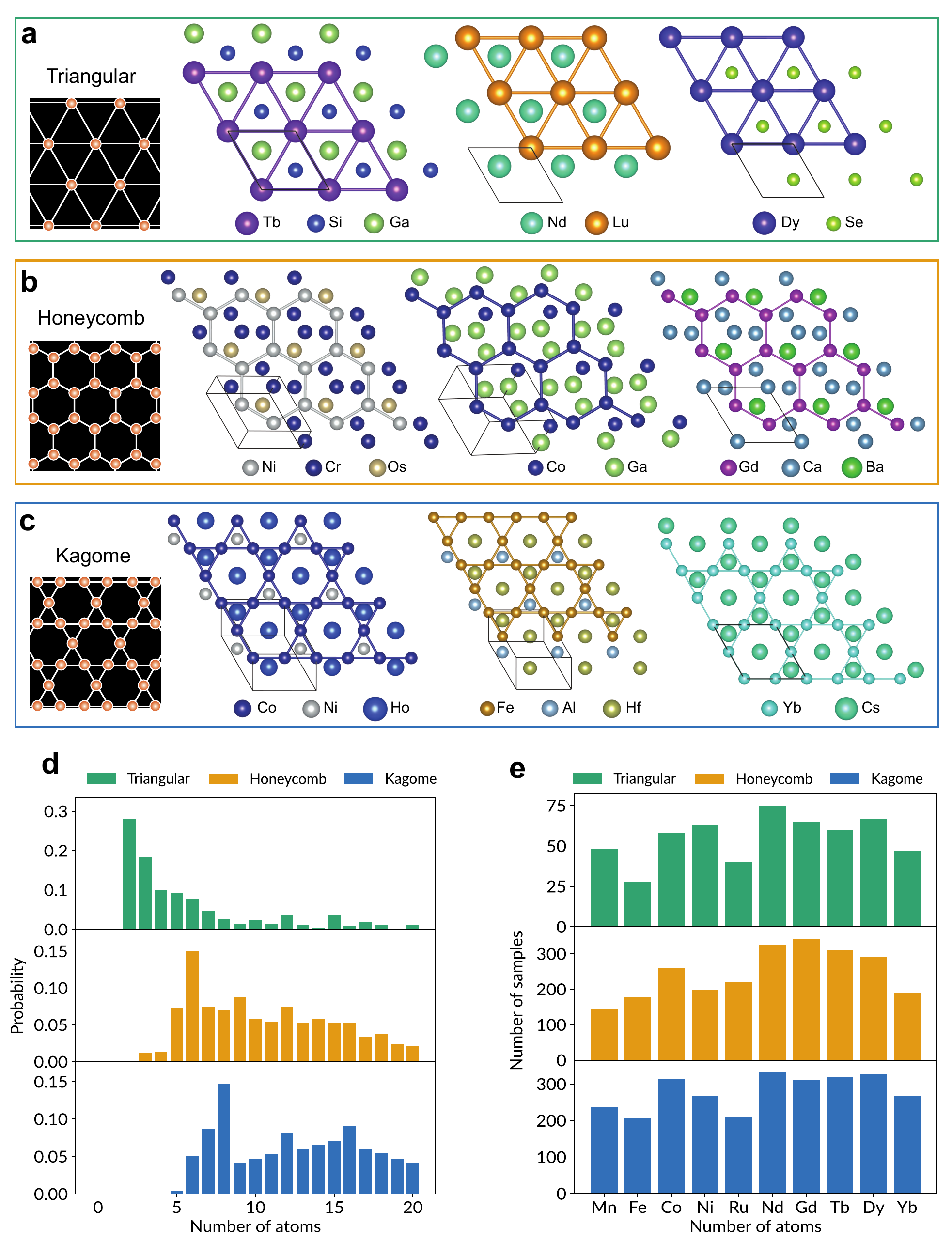}
\centering
\caption{
\textbf{Generated materials with three primary types of archimedean lattices.}    
Archimedean lattice patterns and generated material structures are displayed for \textbf{a.} Triangular, \textbf{b.} Honeycomb, and \textbf{c.} Kagome lattices.  \textbf{d.} The sampling profile of the number of atoms per unit cell $N$, generated by measuring the survival ratio from a uniform sampling of $N$. \textbf{e.}  The number of materials remaining after pre-screening is presented for the common magnetic atom types in each of the primary geometrical patterns. 
}
\label{gen_out}
\end{figure}

Figures \ref{gen_out}a-c display the results of materials generation constrained by three primary AL types: (a) triangular, (b) honeycomb, and (c) kagome. The AL structures are formed in the generated structures, as SCIGEN algorithm has guided them to be formed as pre-defined. The positions of the other unconstrained atoms are not specified rigorously but are often found to reside on the sites that bridge the AL atoms. For triangular lattices (Fig. \ref{gen_out}a), each of the unconstrained atoms is placed on the sites connecting with three magnetic atoms forming an equilateral triangle. For honeycomb lattices (Fig. \ref{gen_out}b), we often observe materials with one unconstrained atom at the center of the hexagon formed with magnetic atoms within the same plane. As to kagome materials (Fig. \ref{gen_out}c), unconstrained atoms bridge the equilateral triangles and the hexagons of kagome lattice layers. If the space inside the polygons is too small compared to the atomic radii, the filler atoms will be pushed outside the AL plane. On the other hand, large polygons like hexagons can accommodate the filler atoms to  fit within the same plane. 

To generate constrained material structures with a higher likelihood of stability, we develop a scheme for sampling initial conditions. We analyze the ratio of stable outputs, defined as the survival ratio after the multi-staged pre-screening processes. First, we sample the number of atoms per unit cell ($N$) from a uniform distribution to identify which $N$ values are more likely to pass the stability pre-screening. This results in a probability distribution $p_N$ values based on their pre-screened stability. We then use this probability distribution $p_N$ to sample $N$ for initializing the large-scale generative process. Figure \ref{gen_out}d shows the sampling profile of $N$, which covers all of the 10 common magnetic atoms as the vertices of AL structures. For triangular lattice materials, smaller $N$ values show higher success rates, whereas larger $N$ values are favored for honeycomb and kagome. This result is reasonable since the AL type is directly linked to the unit cell size, which is a linear function of bond lengths for each class of AL types. For triangular lattice, the lattice parameters $l_{1}$ and $l_{2}$ are the same as the bond length of the neighbor node, while for honeycomb and kagome lattices, the lattice parameters are $\sqrt{3}$ and 2 times of the bond length, respectively. In the case where many atoms are packed into the unit cell of a small cross-section of the AL, the cell needs to be ``tall'', i.e., $l_3$ needs to be larger with respect to  $l_{1}$ and $l_{2}$. Next, we survey which magnetic atoms are suitable as the vertices of ALs. Figure \ref{gen_out}e presents the number of stable materials after the prescreening with respect to magnetic atom types, which we analyze from the set of 3000 generated materials for each lattice type and each magnetic atom. Despite variations, all magnetic atoms are shown to be able to form AL structures. Therefore, we choose to sample atom types for AL vertices with equal probabilities for large-scale materials generation and database construction. Methods section and Supplementary Information 2 describe in detail the sampling schemes for the initialization conditions.

\begin{figure}[ht!]
\includegraphics[width=1.0\textwidth]{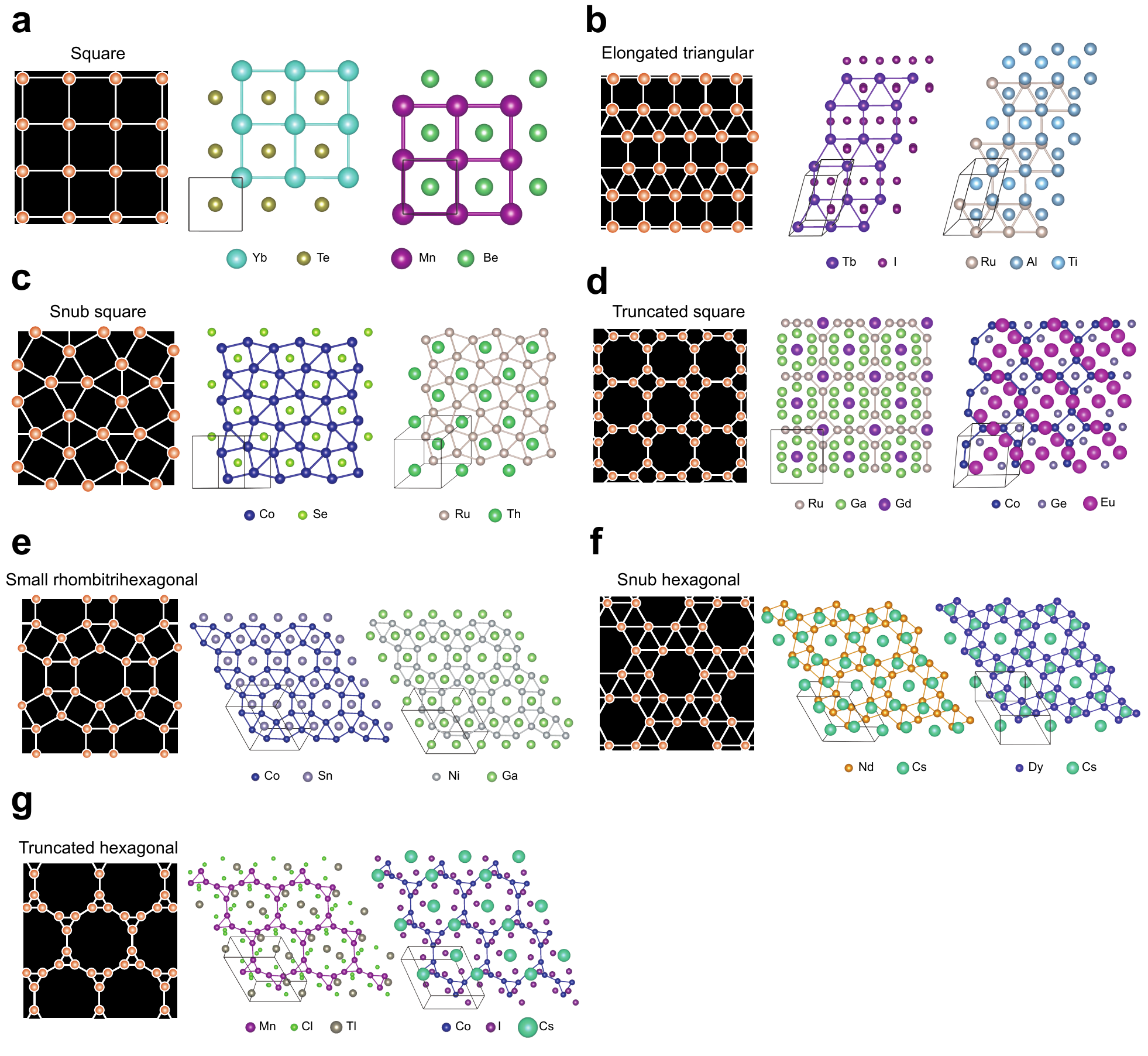}
\centering
\caption{
\textbf{Generated materials with other Archimedean lattice structures.}   
Materials examples covering the rest of Archimedean lattices are presented, with \textbf{a.} Square \textbf{b.} Elongated triangular \textbf{c.} Snub square \textbf{d.} Truncated square \textbf{e.} Small rhombitrihexagonal \textbf{f.} Snub hexagonal \textbf{g.} Truncated hexagonal. In each subplot, the AL pattern and two examples of generated materials are displayed.  
}

\label{more_arch}
\end{figure}

Our exploration of materials with geometrical constraints does not end with the three primary types of ALs but can apply to other geometrical patterns. In contrast to the common types of triangular, honeycomb, and kagome lattices, magnetic systems known to fit in other ALs are extremely rare. Figure \ref{more_arch} showcases 3$\times$3$\times$1 supercells of the generated materials with seven other types of Archimedean lattices: Square, Elongated triangular, Snub square, Truncated square, Small rhombitrihexagonal, Snub hexagonal, and Truncated hexagonal. One type of AL lattice, Great rhombitrihexagonal, is not presented due to the challenge to generate stable materials. The unconstrained atoms within these materials often play a critical role in the overall stability of the structures. They tend to bridge gaps between structured lattice layers, either by sitting at the center of polygons on the same plane or contacting all vertices of the polygon structures, effectively stabilizing the AL layers. This bridging is not just a passive consequence of the material generation process but actively contributes to the mechanical and thermal stability of materials\cite{zimmermann2020local}. Interestingly, even when not explicitly constrained to form specific lattice structures, these unconstrained atoms frequently organize into recognizable Archimedean patterns. This trend could suggest an inherent preference or stability in the configurations of AL whose vertices are equivalent with respect to the local coordinates.

\subsection*{Materials generation with Lieb-like lattice structures}
\begin{figure}[ht!]
\includegraphics[width=1.0\textwidth]{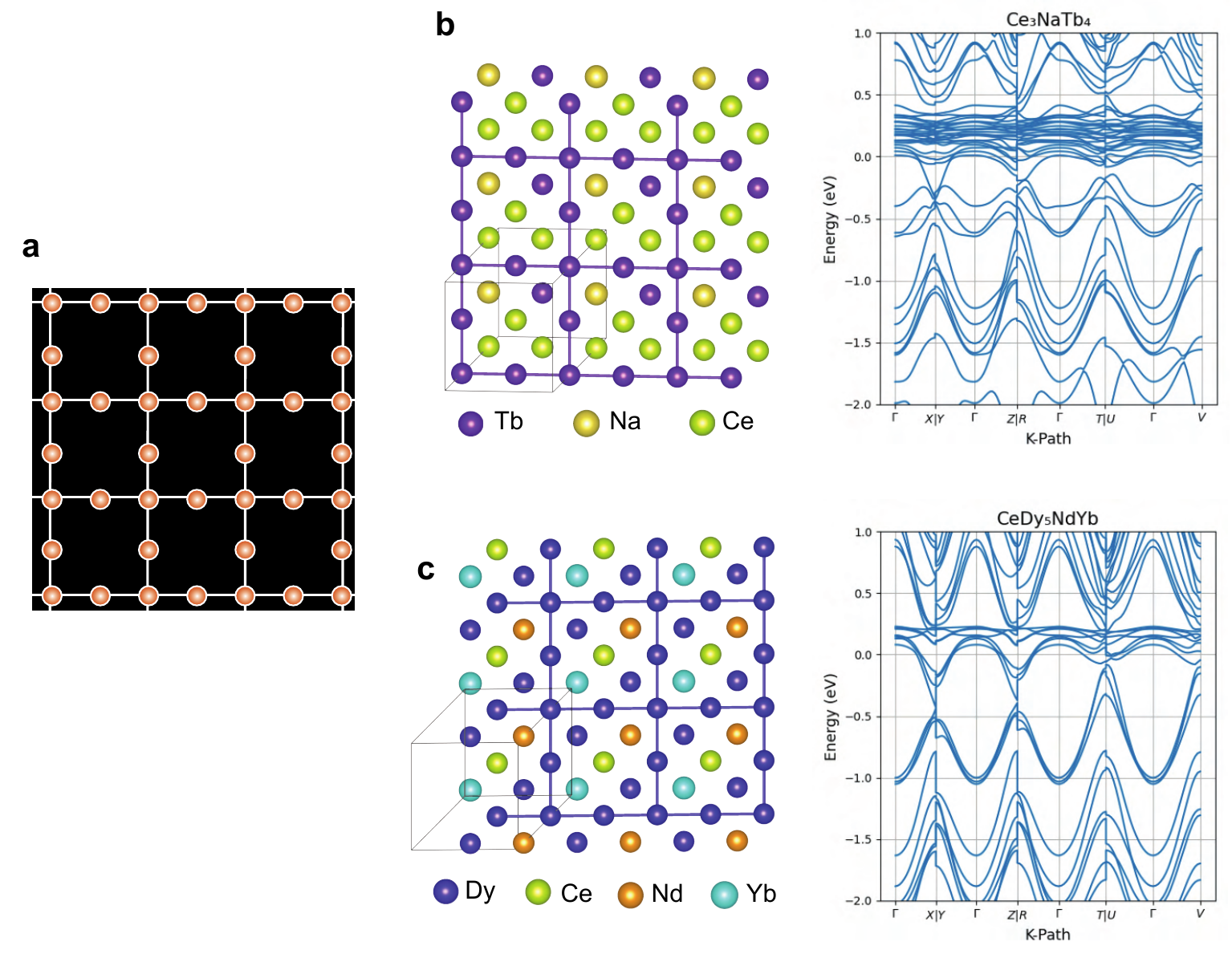}
\centering
\caption{
\textbf{Generated materials of a Lieb-like lattice.}     \textbf{a.} The Lieb lattice pattern that we integrate into the generated structures. The supercell of the Lieb-like lattice materials and the flat band structures of \textbf{b.} 
 \ce{Ce3NaTb4} and \textbf{c.} \ce{CeDy3NdYb}. We plot the band structures by setting the Fermi level $E_F$ to 0 eV, and the flat bands in both examples are slightly (0.1 -- 0.2 eV) above the Fermi level. 
}

\label{lieb}
\end{figure}

The Lieb lattice is a variation of AL that consists of a square lattice with additional atoms located at the centers of each edge of the squares, as visualized in Fig. \ref{lieb}a. Each unit cell of the Lieb lattice contains three atoms. The geometry of the Lieb lattice can lead to magnetic frustration when interacting spins are placed at each lattice site, as we expect for AL structures. This can result in complex magnetic states, which are of significant interest for studying quantum magnetism. Beyond that, Lieb lattices are studied to possess characteristic electronic properties. One key feature of the Lieb lattice is the presence of a flat electronic band\cite{slot2017experimental}. Contrary to localized atomic orbitals, the flat bands formed from the Lieb lattice originate from the destructive quantum interference effect which quenches the kinetic energy. This may lead to interesting physical phenomena such as enhanced electron correlation and high-temperature superconductivity\cite{yin2022topological, kang2020topological}. Also, recent research has shown that the Lieb lattice can exhibit non-trivial topological properties when subjected to various perturbations\cite{tsai2015interaction}. However, the Lieb lattice has mainly been achieved in artificial systems like photonic crystals\cite{mukherjee2015observation, vicencio2015observation}, and atomic solids that can host Lieb lattice are extremely rare.  

In this work, we also focus on the Lieb-like lattice where magnetic atoms sit on the Lieb lattice. Figures \ref{lieb}b,c show the generated materials with Lieb-lattice-based crystal structures and their calculated band structures. In these generated materials, magnetic atoms such as terbium (Tb) and dysprosium (Dy) are strategically positioned at the nodes of the Lieb lattice. Following structural relaxation through DFT calculations, the integrity of the Lieb lattice architecture is maintained, and the structures exhibit the anticipated flat-band characteristics close to the Fermi level. These outcomes demonstrate SCIGEN's ability to generate new, stable materials with exotic geometric patterns, even when there are very few known materials that fit the desired geometric pattern.

\subsection*{Database of the materials with Archimedean lattice}
As detailed in Supplementary Information 7, we generate an AL materials database using SCIGEN. The database contains three components: the total 7.87 million materials generated by the SCIGEN model, the 790 thousand materials that survived four stages of stability pre-screening processes, and 24,743 out of 26,000 sub-sampled materials in which DFT calculations are successfully converged. By systematically extending our exploration to encompass a wider range of ALs, we can investigate new exotic magnetic orderings, discover porous structures beyond zeolites, and explore new electronic flatband structures, among other possibilities.

\section*{Conclusions}

In this work, we present SCIGEN, a new generative model aimed at discovering quantum material candidates that adhere to geometric constraints. Our method leverages an AL layer within the crystal structure to identify potential quantum materials. These materials have been validated through DFT to ensure that their relaxed structures are consistent with the machine-learning generations.  

To further enhance the validity of SCIGEN, it is crucial to conduct experimental verification through the synthesis of these machine-generated materials. Computation-aided synthesizability check may involve additional analysis in binary, ternary, or other more complex phase diagrams. Setting aside experimental validation, SCIGEN paves an avenue for a few future directions. By focusing on atomic arrangements, we can explore additional geometry-related constraints, such as bonding types, coordination numbers, short-range orderings, and point-group and space group symmetries, during materials generation. Additionally, integrating other diffusion channels, like through the virtual node approach, allows us to incorporate more complex constraints such as defect constraints and magnetic interaction constraints, broadening the scope of the SCIGEN model. Moreover, conditioning the generation process with targeted functionalities, such as specific electrical and optoelectronic properties, or sustainability or environmental impact of materials, can lead to the direct creation of materials with tailored performance. Our SCIGEN model represents a general machine learning-based framework for discovering new quantum materials. It leverages information typically absent from crystal structure databases, offering deeper insights into the structure-property relationships of emerging quantum materials.\\

\section*{Methods}

\subsection*{Initialization of the Archimedean Lattices}
We describe the workflow to initialize the materials generation process related to the both AL and the entire structure of crystal for the diffusion model. This initialization process involves a few steps: the choice of AL, the atom types, and the total number of atoms per unit cell. Here we provide a summary of the initialization process in Fig. \ref{gen_out}b, where more detailed scheme can be found in Supplementary Information 2. 

First, we assign the required geometric domain condition to the crystals. In SCIGEN, we specify one of the AL structures, such as triangular, honeycomb, or kagome lattice, as geometric domain condition. Each type of AL requires the number of vertices per unit cell and the size of the unit cell. Supplementary Information 1 presents the geometric patterns and the preliminary profiles of all AL and Lieb lattices.

Second, we choose the constrained atom type $\mathcal{A}^c$ placed on the vertices of the AL structure assigned above. To generate candidate materials which may host geometrically frustrated quantum magnetism, we specify 10 types of common magnetic atoms (Mn, Fe, Co, Ni, Ru, Nd, Gd, Tb, Dy, Yb) on the vertices. The atom types are chosen independently from the AL choice above. 

Third, we sample the constrained magnetic bond lengths $d^c$, aka the distances between the nearest-neighbor magnetic atoms forming the ALs. For each magnetic atom type $\mathcal{A}^c$, we generated the profile of the bond lengths by sampling the nearest-neighbor distances between the corresponding atoms in the MP-20 dataset\cite{xie2018crystal, jain2013commentary} using CrystalNN\cite{pan2021benchmarking}. To ensure the nearest-neighbor distances do not become significantly close, we cut the minimum lengths by the metallic radii\cite{zachariasen1973metallic} for each atom type. The bond length distribution for each magnetic atom type $\mathcal{A}^c$, $p_{d^c}(\mathcal{A}^c)$, is presented in Supplementary Information 2. 

Finally, we sample the total number of atoms per unit cell $N$. Each of the ALs has the distribution of the preferable $N$ values with better stability. We generate $p_N$, the stable materials probability distribution of $N$. We can sample $N$ from $p_N$ as the sampling profile of $N$ for each AL type. The sampling profile of both $p_N$ and $p_N(\mathcal{A}^c)$, which is distribution of with each magnetic atom type $\mathcal{A}^c$, are displayed in Supplementary Information 2.

To impose Archimedean lattice as the constraints, we organize masks $\bs{m} = (\bs{m}^{\bs{L}}, \bs{m}^{\bs{F}}, \bs{m}^{\bs{A}})$, which give constraints to the lattice $\bs{L}$, fractional coordinates $\bs{F}$, and atom types $\bs{A}$ respectively. $\bs{m}^{\bs{L}}$ is equal to 1 for the two lattice basis vectors $\bs{l}_{1}$ and $\bs{l}_{2}$, defining the unit cell of AL layer plane. $\bs{m}^{\bs{L}}$ is equal to 0 for $\bs{l}_{3}$, as we let the diffusion model generate $\bs{l}_{3}$ without explicit constraints. We assign $\bs{m}^{\bs{F}}$ is equal to 1 for the $i$-th atoms ($i\in [1, N^c]$) to guide them to be placed at the vertex positions of AL layers. The same rule applies for $\bs{m}^{\bs{A}}$ so that the atoms at AL vertices result in the magnetic atom types $\mathcal{A}^c$. 

\subsection*{Integration of constrained and unconstrained components to guide materials generation}

We design SCIGEN as a generic framework applicable for any diffusion model as a base model. Without loss of generality, let the pre-trained base model represent a periodic structure as $\bs{M}_0$, with $T$ diffusion steps, a sampling probability prior $P_T$, diffusion inference model $q$, which is normally chosen to map the materials distribution of the training dataset to $P_T$, and denoising generative model $p$ which needs to be trained. The diffusion inference model $q$ works by iteratively injecting noise to the input structure, $\bs{M}_0$. The inference probability of most diffusion models is a Markov process, i.e., the probability of diffusing $\bs{M}_0$ for $t$ steps to $\bs{M}_t$ can be written as
\begin{equation}
    P(\bs{M}_t|\bs{M}_0) = q_{0, t}(\bs{M}_t|\bs{M}_0) = \prod_{s=1}^t q_{s-1, s}(\bs{M}_{s}|\bs{M}_{s-1})
\end{equation}
with
\begin{equation}
    P(\bs{M}_T|\bs{M}_0) = q_{0, T}(\bs{M}_T|\bs{M}_0) \approx P_T.
\end{equation}
A well-trained diffusion model should have denoising generative model $p$ that can inverse the diffusion, i.e., for a denoising step from $\bs{M}_t$ to $\bs{M}_{t-1}$,
\begin{equation}
    p_{t, t-1}(\bs{M}_{t-1}|\bs{M}_{t}) \approx q_{t, t-1}(\bs{M}_{t-1}|\bs{M}_{t}).
\end{equation}
Here, the subscripts of $p$ and $q$ indicate the initial and final time steps that the models are applied to, e.g., $q_{t_1, t_2}$ is the diffusion inference from time step $t_1$ to $t_2$. Note that, since $q$ is normally chosen to be a simple probabilistic function, we cannot easily find its inverse. Hence, the training for the denoising generative model is required.

Our approach to material design is summarized in Algorithm \ref{alg_matinp}, where integration of geometrical constraints plays a pivotal role in ensuring that certain structural elements, like lattice configurations or specific atomic distributions, adhere closely to predefined parameters. This method effectively blends prescribed structural characteristics with the creative latitude allowed in other aspects of the material's architecture. Previously, a generative model with unmasked areas as constraints has been employed in image generation for image inpainting, such as RePaint method \cite{lugmayr2022repaint}. However, the geometrical pattern constraint for crystal generation is still challenging and differs in a few ways. The generation of crystal generative model with geometrical constraint involves several key steps:
\begin{enumerate}
    \item \textbf{Adding noise to the constraint structures:} Initially, we introduce noise to a structure which is randomly selected from structures that satisfy the target constraints (This constrained structure can be unstable, or unrealistic as long as it contains the target constraints.) with the diffusion inference model, $q$, to get diffused constrained structures for each time step $t \in [1, T]$. This operation is aimed at creating a predefined pathway for denoising the constrained structures. The unconstrained components of the crystals are guided by this known denoising pathway, which results in the presence of constrained components in the final outputs.
    \item \noindent\textbf{Denoising the unconstrained structures with a base diffusion model:} Concurrently, the parts of the structure that are unconstrained from these specific constraints undergo a normal denoising process with $p$. This process, facilitated by the base model, iteratively refines these regions by methodically reducing the introduced noise, thereby nudging them toward physically realistic configurations.
    \item \noindent\textbf{Integration of the constrained and unconstrained structures:} For each denoising step, after processing both parts independently, they are carefully recombined. This combination is critical as it ensures the integrity of the predefined constraints is maintained while integrating seamlessly with the freely generated segments. This method preserves essential structural features and fosters innovation in material design.
\end{enumerate}

Algorithm \ref{alg_matinp} presents the SCIGEN sampling procedure designed to generate material structures with structural constraints. The algorithm utilizes a diffusion model to iteratively refine structures, ensuring the generated structures contain specific geometry as constraints. The procedure begins with the initialization of constrained structures, indicated with superscript $c$, $\bs{M}^{c}_0$, along with the corresponding masks $\bs{m}$, which indicate the constrained components in $\bs{M}^{c}_0$ with binary masking, i.e., assigns value of 1 to constrained, and 0 to the unconstrained components. The final-time-step unconstrained structure, indicated with superscript $u$, $\bs{M}^u_T$, and constrained structure $\bs{M}^c_T$ are sampled from the probability prior $P_T$ of the base model. Through masking, we obtain the final-time-step structure $\bs{M}_T$ that contains the constrained components from $\bs{M}^c_T$, and the remaining parts from $\bs{M}^u_T$ formulated as $\bs{M}_T \gets \bs{m} \odot \bs{M}^{c}_T + (1-\bs{m}) \odot \bs{M}^{u}_T$ where $\odot$ represents a component-wise multiplication. Basically, the components of $\bs{M}_T$ that got masked (values in $\bs{m}$ equal to 1) come from $\bs{M}^{c}_T$ while the remaining components (values in $\bs{m}$ equal to 0) come from $\bs{M}^{u}_T$

The iterative process begins from the final time step $T$ and proceeds backward to 0. For each time step $t$, the structure $\bs{M}_t$ undergoes the denoising process giving the distribution of the unconstrained structure at time step $t-1$, $\bs{M}^u_{t-1}$, as $p_{t, t-1}(\bs{M}^u_{t-1}|\bs{M}_t)$. Concurrently, the diffusion process gives the distribution of the constrained structure $\bs{M}^c_{t-1}$ as $q_{0, t-1}(\bs{M}^c_{t-1}|\bs{M}^c_{0})$. Then, the unconstrained structure $\bs{M}^u_{t-1}$, and constrained structure $\bs{M}^c_{t-1}$ are sampled from their corresponding probability distributions. The structure $\bs{M}_t$ is updated by combining the sampled constrained and unconstrained parts using the mask $\bs{m}$ similar to the final-time-step case. The process continues iteratively until the initial time step is reached, at which point the refined structure $\bs{M}_0$ is returned. This ensures that the generated material structures respect the given constraints and exhibit realistic and viable configurations. Supplementary Information 3 provides the schematic explanation of the denoising process, as well as the integration of constrained and unconstrained components of material structures. 

This approach effectively integrates structural constraints into the diffusion model, enabling the generation of novel material structures that align with the AL structures as the predefined requirements. Using masks to combine constrained and unconstrained parts ensures that the constraints are maintained throughout the iterative refinement process, resulting in high-quality material structures suitable for practical applications.

\begin{algorithm}
    \caption{Structural Constraint Integration in Material Generation Procedure}
    \begin{algorithmic}[1]
        \State \textbf{Input}: constrained structure $\bs{M}^c_0$, constraint mask $\bs{m}$, diffusion inference model $q$, denoising generative model $p$, number of steps $T$, probability prior $P_T$
        \State Sample $\bs{M}^{u}_T \sim P_T$, $\bs{M}^{c}_T \sim P_T$
        \State $\bs{M}_T\gets \bs{m} \odot \bs{M}^{c}_T + (1-\bs{m}) \odot \bs{M}^{u}_T$
        \For{$t = T, \dots, 1$}
            \State Sample $\bs{M}^c_{t-1}\sim q_{0, t-1}(\bs{M}^c_{t-1}|\bs{M}^c_0)$
            \State Sample $\bs{M}^u_{t-1}\sim p_{t,t-1}(\bs{M}^u_{t-1}|\bs{M}_{t})$
            \State $\bs{M}_{t-1} \gets \bs{m} \odot \bs{M}^{c}_{t-1} + (1-\bs{m}) \odot \bs{M}^{u}_{t-1}$
        \EndFor
        \State \Return $\bs{M}_0$.
    \end{algorithmic}
\label{alg_matinp}
\end{algorithm}

To demonstrate the algorithm for the generation of AL materials, we chose DiffCSP\cite{jiao2024crystal} as the base model of SCIGEN before applying geometric constraint. In DiffCSP, the structure representation got divided into three components $\bs{M}=(\bs{L}, \bs{F}, \bs{A})$: the lattice matrix containing three basis vectors $\bs{L}=\left[\bs{l}_{1}, \bs{l}_{2}, \bs{l}_{3}\right] \in \mathbb{R}^{3 \times 3}$, the fractional coordinates $\bs{F}=\left[\bs{f}_{1}, \bs{f}_{2}, \ldots, \bs{f}_{N}\right] \in [0, 1)^{3 \times N}$, and one-hot representations of atom types $\bs{A}=\left[\bs{a}_{1}, \bs{a}_{2}, \ldots, \bs{a}_{N}\right] \in$ $[0, 1]^{h \times N}$. Using these components, the infinite periodic crystal can be described as 
\begin{equation}
\left\{\left(\bs{a}_{i}, \bs{x}_{i}\right) \mid \bs{x}_{i} = \bs{L}\cdot (\bs{f}_i+\bs{k}), \forall \bs{k} \in \mathbb{Z}^{3 \times 1}, \forall i\in [1..N]\right\}
\label{eq_period_crystal}
\end{equation}
which tell all atomic types $\bs{a}$, and Cartesian coordinates $\bs{x}$ of every atoms in the structure. DiffCSP uses normalized Gaussian diffusion in its diffusion inference model $q$ which have standard normal distribution as probability prior, i.e., $P_T=\mathcal{N}(0, \bs{\textit{I}})$. Furthermore, the diffusion is applied independently between components, making it possible to split the model as $q=(q^{\bs{L}}, q^{\bs{F}}, q^{\bs{A}})$ where the superscripts indicate the components that the diffusion acts on. We can write the split diffusion inference model of DiffCSP as
\begin{equation}
    q_{t, t+1}(\bs{M}_{t+1}|\bs{M}_{t}) = q^{\bs{L}}_{t, t+1}(\bs{L}_{t+1}|\bs{L}_{t})\cdot q^{\bs{F}}_{t, t+1}(\bs{F}_{t+1}|\bs{F}_{t})\cdot q^{\bs{A}}_{t, t+1}(\bs{A}_{t+1}|\bs{A}_{t}).
\end{equation}

\begin{algorithm}
    \caption{Structural Constraint Integration in Material Generation with DiffCSP}
    \begin{algorithmic}[1]
        \State \textbf{Input}: constrained structure $\bs{M}^c_0=(\bs{L}^c_0, \bs{F}^c_0, \bs{A}^c_0)$, constraint mask $\bs{m}=(\bs{m}^{\bs{L}}, \bs{m}^{\bs{F}}, \bs{m}^{\bs{A}})$, diffusion inference model $q=(q^{\bs{L}}, q^{\bs{F}}, q^{\bs{A}})$, denoising generative model $p=(p^{\bs{L}}, p^{\bs{F},p}, p^{\bs{F},c}, p^{\bs{A}})$, number of steps $T$
        \State Sample $\bs{M}^{u}_T=(\bs{L}^u_T, \bs{F}^u_T, \bs{A}^u_T)\sim \mathcal{N}(0, \bs{\textit{I}})$, $\bs{M}^{c}_T=(\bs{L}^c_T, \bs{F}^c_T, \bs{A}^c_T) \sim \mathcal{N}(0, \bs{\textit{I}})$
        \State $\bs{M}_T=(\bs{L}_T, \bs{F}_T, \bs{A}_T)  \gets \bs{m} \odot \bs{M}^{c}_T + (1-\bs{m}) \odot \bs{M}^{u}_T$
        \For{$t = T, \dots, 1$}
            \State Sample $\bs{M}^c_{t-1}=(\bs{L}^c_{t-1}, \bs{F}^c_{t-1}, \bs{A}^c_{t-1}) \sim q_{0, t-1}(\bs{M}^c_{t-1}|\bs{M}^c_0)$
            \State Sample $\bs{L}^u_{t-1}\sim p^{\bs{L}}_{t,t-1}(\bs{L}^u_{t-1}|\bs{M}_{t})$, $\bs{A}^u_{t-1}\sim p^{\bs{A}}_{t,t-1}(\bs{A}^u_{t-1}|\bs{M}_{t})$
            \State $\bs{L}_{t-1} \gets \bs{m}^{\bs{L}} \odot \bs{L}^{c}_{t-1} + (1-\bs{m}^{\bs{L}}) \odot \bs{L}^{u}_{t-1}$
            \State $\bs{A}_{t-1} \gets \bs{m}^{\bs{A}} \odot \bs{A}^{c}_{t-1} + (1-\bs{m}^{\bs{A}}) \odot \bs{A}^{u}_{t-1}$
            \State Sample $\bs{F}^c_{t-\frac{1}{2}} \sim q^{\bs{F}}_{0, t-1}(\bs{F}^c_{t-1}|\bs{F}^c_0)$
            \State Sample $\bs{F}^u_{t-\frac{1}{2}}\sim p^{\bs{F},p}_{t,t-1}(\bs{F}^u_{t-\frac{1}{2}}|\bs{M}_{t})$
            \State $\bs{F}_{t-\frac{1}{2}} \gets \bs{m}^{\bs{F}} \odot \bs{F}^c_{t-\frac{1}{2}} + (1-\bs{m}^{\bs{F}}) \odot \bs{F}^u_{t-\frac{1}{2}}$
            \State Sample $\bs{F}^u_{t-1}\sim p^{\bs{F},c}_{t,t-1}(\bs{F}^u_{t-1}|\bs{L}_{t-1}, \bs{F}_{t-\frac{1}{2}}, \bs{A}_{t-1})$
            \State $\bs{F}_{t-1} \gets \bs{m}^{\bs{F}} \odot \bs{F}^c_{t-1} + (1-\bs{m}^{\bs{F}}) \odot \bs{F}^u_{t-1}$
        \EndFor
        \State \Return $\bs{M}_0=(\bs{L}_0, \bs{F}_0, \bs{A}_0)$.
    \end{algorithmic}
\label{alg_matinp_diffcsp}
\end{algorithm}

Since the denoising generative process of DiffCSP utilizes Predictor-Corrector sampling\cite{song2020score} mechanism on the fractional coordinate components only, the model need to be split into $p=(p^{\bs{L}}, p^{\bs{F},p}, p^{\bs{F},c}, p^{\bs{A}})$ where the boldface superscripts indicates the components that the diffusion gives while $p$ and $c$ superscripts indicate predictor and corrector sub-models, respectively. We can write the split denoising generative model of DiffCSP as
\begin{equation}
    p_{t, t-1}(\bs{M}_{t-1}|\bs{M}_{t}) = p^{\bs{L}}_{t, t-1}(\bs{L}_{t-1}|\bs{M}_{t})\cdot p^{\bs{F},p}_{t, t-1}(\bs{F}_{t-\frac{1}{2}}|\bs{M}_{t})\cdot 
    p^{\bs{F},c}_{t, t-1}(\bs{F}_{t-1}|\bs{L}_{t-1}, \bs{F}_{t-\frac{1}{2}}, \bs{A}_{t-1})\cdot p^{\bs{A}}_{t, t-1}(\bs{A}_{t-1}|\bs{M}_{t}).
\end{equation}
Basically, the lattice, and atomic type components got denoised by their respective model to get $\bs{L}_{t-1}$, and $\bs{A}_{t-1}$, respectively. For the fractional coordinate components, the predictor denoises them to the half-time-step point $\bs{F}_{t-\frac{1}{2}}$, and the corrector uses $\bs{F}_{t-\frac{1}{2}}$, $\bs{L}_{t-1}$, and $\bs{A}_{t-1}$ to predict $\bs{F}_{t-1}$. Because of this additional prediction of $\bs{F}_{t-\frac{1}{2}}$, algorithm \ref{alg_matinp} need to be slightly modified to accommodate the constraints that are also imposed on the $\bs{F}_{t-\frac{1}{2}}$ as shown in algorithm \ref{alg_matinp_diffcsp}. 

\subsection*{Pre-screening procedure to retrieve stable materials structures}
Following the generation of materials constrained by AL structures, it becomes essential to evaluate their stability. Due to the high volume of generated candidates—often reaching into the millions—a rapid yet reliable method is necessary to assess stability. Here, we describe our four-staged pre-screening of materials based on a series of stability criteria.

\subsubsection*{Charge Neutrality}
Materials must be electrically neutral to ensure stability and real-world applicability. We employed the SMACT approach\cite{davies2019smact} to evaluate the charge neutrality of generated materials. This process, inspired by methodologies described in the CDVAE approach\cite{xie2021crystal}, ensures that only chemically feasible materials are considered in subsequent steps.

\subsubsection*{Density and Space Occupancy Ratio}
Some generated materials feature densely packed atomic configurations, which are unrealistic in actual crystalline phases. To address this, we compare these materials against a reference dataset (MP-20) to identify and eliminate those with excessively high atom densities. The space occupancy ratio $R_{occ}$ is calculated as
\begin{equation}
R_{occ} = \frac{\sum^{N}_{i=1} \frac{4\pi r_i^3}{3}}{V_{cell}}
\end{equation}

\noindent where $r_i$ is the radius of the $i$-th atom and $V_{cell}$ is the volume of the unit cell. Materials in MP-20 observed a similar distribution of $R_{occ}$ that is independent of $N$, as we can find in Supplementary Information 6. We discard materials with an $R_{occ}$ value exceeding 1.7, a threshold based on the distribution of $R_{occ}$ in the MP-20 dataset.

\subsubsection*{Graph Neural Network Classifiers for Stability Evaluation}
To rapidly assess the stability of the remaining material candidates, we utilize graph neural networks (GNNs) based on the E3NN\cite{geiger2022e3nn, chen2021direct} framework, designed for their efficiency in handling crystallographic data. We develop two models:

\begin{itemize}
    \item \textbf{GNN for Stability classification $\Psi_1$:} This model predicts whether a material's energy above the convex hull ($E_{\text{hull}}$) is below a threshold of 0.1 eV, which is indicative of thermodynamic stability. The model is trained using data from the Matbench-discovery\cite{riebesell2023matbench}.

    \item \textbf{GNN for Stability classification $\Psi_2$:} Recognizing the potential for our model to generate materials that diverge from known stable structures, this classifier distinguishes between pristine and diffused structures. Training data includes original structures from the MP-20 dataset and the ones with added Gaussian noise on unit cell matrix $\bs{L}$ and fractional coordinates $\bs{F}$, simulating potential inaccuracies in atomic positions during generation. As the training dataset of  $\Psi_2$, we diffuse $\bs{L}$ and $\bs{F}$ as $\bs{L}' = \bs{L} + (l_1, l_2, l_3)^T \cdot \mathcal{N}(0,\sigma_d^2\bs{\textit{I}})$ and $\bs{F}' = w(\bs{F} + \mathcal{N}(0, \sigma_d^2\bs{\textit{I}}))$, respectively. Here, $\sigma_d$ regulates the level of diffusion, and $w(\cdot)$ is a wrapping function that adjust fractional coordinates to fit within $[0, 1)$. We decide that 1$\%$ diffusion materials ($\sigma_d=0.01$) are stable, but 5$\%$ diffusion materials ($\sigma_d=0.05$) are unstable. This model helps ensure that even stable materials are not overly distorted or unrealistic.
\end{itemize}

The two GNN-based classifiers are in simple architecture, but presented high accuracy. We present the confusion matrices of the two models in Supplementary Information 6. The prediction accuracy for the test dataset is 0.83 and 0.99, respectively. 

The four-staged pre-screening filters could provide us with an extremely efficient screening tool to evaluate the stability of the generated materials. We argue this stability evaluation is valid, as we observe the materials gain stability as it goes through the denoising process from the noisy structures ($\bs{M}_{T}$) to the pristine ones ($\bs{M}_{0}$). Moreover, more than 50\% of pre-screened materials survive DFT relaxation, indicating the efficacy of the pre-screening process. The detail of the stability evaluation is shown in Supplementary Information 6. 

\subsection*{DFT for stability evaluation and structural relaxation}
The candidate models are further evaluated with DFT for potential structural stability. Due to the high cost of DFT calculations, it is necessary to balance the accuracy and throughput. In the first stage of DFT screening, we choose to use a relatively coarse treatment of the electronic structure to evaluate as many candidates as possible (up to 26,000). Planewave DFT calculations are performed using the Vienna Ab initio Simulation Package (VASP)\cite{kresse1996efficient}. The calculation used Projector Augmented Wave (PAW) method\cite{blochl1994projector, kresse1999ultrasoft} to describe the effects of core electrons and Perdew-Burke-Ernzerhof (PBE)\cite{perdew1996generalized} implementation of the Generalized Gradient Approximation (GGA) for the exchange-correlation functional. The energy cutoff is $1.2\cdot\text{max(ENMAX)}$ for the plane-wave basis of the valence electrons. The electronic structure is calculated on $\Gamma$-centered mesh for the unit cell (the grid length density is 5 $k$-points per nm$^{-1}$). The total energy tolerance for electronic energy minimization is $10^{-6}$ eV, and the energy criterion for structure optimization is $10^{-5}$ eV. The maximum number of steps is 60 for electronic self-consistent calculation and 150 for structural optimization. During the structural relaxation, the symmetry of the crystal is maintained while the cell shape/size and all atomic coordinates are allowed to relax. Non-spin-polarized calculation is employed for this initial screening. A small fraction of models failed the electronic structure calculation, which are terminated and the corresponding candidates are considered unstable. For the candidates with completed VASP calculation (either because the energy criterion is reached or the maximum number of relaxation steps is reached), the following quantities are extracted/calculated as indicators of potential stability: (1) maximum interatomic force after relaxation, (2) initial and final total energy, (3) average changes in lattice constants, (4) average changes in atomic coordinates. These quantities are then analyzed and compared to identify potentially stable candidates worth further and more rigorous evaluations.

\subsection*{Band structure calculation for Lieb-like lattice materials}
The DFT band structure calculations for Lieb-like lattice structures are performed using VASP. PAW method and PBE exchange-correlation functional are used for all DFT calculations. The initial electronic structure calculations are performed on a K-point mesh centered at Gamma point with resolved value $k_{mesh}=0.03\cdot 2\pi$/$\mathring{\text{A}}$ for each structure. The band structure is subsequently calculated on a high symmetry path generated by the VASPKIT code\cite{wang2021vaspkit}.

\subsection*{Data visualization}
We use VESTA \cite{momma2011vesta} to visualize the materials structures presented in the main article. For Supplementary Information, we utilize OVITO \cite{stukowski2009visualization} to visualize the materials structures. 

\section*{Data Availability Statement} 
We compile a comprehensive database of AL materials generated by SCIGEN. The dataset provides the folders of the entire generated materials (7.87 million), the survived materials after the four-staged pre-screening process (790 thousand materials), and DFT-relaxed structures (24,743). The folder with DFT calculation contains materials structures before and after relaxation. The Supplementary dataset is available in Figshare repository\cite{figshare}.

\section*{Code Availability Statement} 
 The source code is available at (\url{https://github.com/RyotaroOKabe/SCIGEN}).

\section*{Acknowledgements}

RO and ML thank C Batista, A Christianson, F Frenkel, A May, R Moore, B Ortiz, and F Ronning for the helpful discussion. RO acknowledges the support from the U.S. Department of Energy (DOE),  Office of Science (SC), Basic Energy Sciences (BES), Award No. DE-SC0021940 and Heiwa Nakajima Foundation. AC acknowledges support from National Science Foundation (NSF) Designing Materials to Revolutionize and Engineer our Future (DMREF) Program with Award No. DMR-2118448. BH and YC are partially supported by the Artificial Intelligence Initiative as part of the Laboratory Directed Research and Development (LDRD) program of Oak Ridge National Laboratory (ORNL), managed by UT-Battelle, LLC, for the US Department of Energy under Contract DE-AC05-00OR22725. Computing resources for a portion of the work were made available through the VirtuES project, funded by the LDRD Program and Compute and Data Environment for Science (CADES) at ORNL. Another portion of simulation results were obtained using the Frontera computing system at the Texas Advanced Computing Center. ML acknowledges the support from NSF ITE-2345084, the Class of 1947 Career Development Chair, and the support from R. Wachnik.


\printbibliography

@Article{Chamorro2021QSLChem,
author={Chamorro, Juan R.
and McQueen, Tyrel M.
and Tran, Thao T.},
title={Chemistry of Quantum Spin Liquids},
journal={Chemical Reviews},
year={2021},
month={3},
day={10},
publisher={American Chemical Society},
volume={121},
number={5},
pages={2898-2934},
issn={0009-2665},
doi={10.1021/acs.chemrev.0c00641},
url={https://doi.org/10.1021/acs.chemrev.0c00641}
}

@article{Savar2017QSL,
doi = {10.1088/0034-4885/80/1/016502},
url = {https://dx.doi.org/10.1088/0034-4885/80/1/016502},
year = {2016},
month = {11},
publisher = {IOP Publishing},
volume = {80},
number = {1},
pages = {016502},
author = {Lucile Savary and Leon Balents},
title = {Quantum spin liquids: a review},
journal = {Reports on Progress in Physics},
}

@Article{Kang2020CoSn,
author={Kang, Mingu
and Fang, Shiang
and Ye, Linda
and Po, Hoi Chun
and Denlinger, Jonathan
and Jozwiak, Chris
and Bostwick, Aaron
and Rotenberg, Eli
and Kaxiras, Efthimios
and Checkelsky, Joseph G.
and Comin, Riccardo},
title={Topological flat bands in frustrated kagome lattice CoSn},
journal={Nature Communications},
year={2020},
month={8},
day={10},
volume={11},
number={1},
pages={4004},
issn={2041-1723},
doi={10.1038/s41467-020-17465-1},
url={https://doi.org/10.1038/s41467-020-17465-1}
}

@article{Hung2024SHG,
doi = {10.1088/1361-6463/ad4a80},
url = {https://dx.doi.org/10.1088/1361-6463/ad4a80},
year = {2024},
month = {5},
publisher = {IOP Publishing},
volume = {57},
number = {33},
pages = {333002},
author = {Nguyen Tuan Hung and Thanh Nguyen and Vuong Van Thanh and Sake Wang and Riichiro Saito and Mingda Li},
title = {Symmetry breaking in 2D materials for optimizing second-harmonic generation},
journal = {Journal of Physics D: Applied Physics},
}

@article{Martin2017FE,
doi = {10.1038/natrevmats.2016.87},
url = {https://www.nature.com/articles/natrevmats201687},
year = {2017},
month = {11},
publisher = {Springer Nature},
volume = {2},
pages = {16087},
author = {Lane Martin and Andrew Rappe },
title = {Thin-film ferroelectric materials and their applications
},
journal = {Nature Reviews Materials },
}

@article{Fu2012TCI,
  title = {Topological Crystalline Insulators},
  author = {Fu, Liang},
  journal = {Phys. Rev. Lett.},
  volume = {106},
  issue = {10},
  pages = {106802},
  numpages = {4},
  year = {2011},
  month = {3},
  publisher = {American Physical Society},
  doi = {10.1103/PhysRevLett.106.106802},
  url = {https://link.aps.org/doi/10.1103/PhysRevLett.106.106802}
}

@article{Armitage2018Weyl,
  title = {Weyl and Dirac semimetals in three-dimensional solids},
  author = {Armitage, N. P. and Mele, E. J. and Vishwanath, Ashvin},
  journal = {Rev. Mod. Phys.},
  volume = {90},
  issue = {1},
  pages = {015001},
  numpages = {57},
  year = {2018},
  month = {1},
  publisher = {American Physical Society},
  doi = {10.1103/RevModPhys.90.015001},
  url = {https://link.aps.org/doi/10.1103/RevModPhys.90.015001}
}

@misc{Bihlmayer2022Rashba,
	author = {},
	title = {{R}ashba-like physics in condensed matter - {N}ature {R}eviews {P}hysics --- nature.com},
	howpublished = {\url{https://www.nature.com/articles/s42254-022-00490-y}},
	year = {},
	note = {[Accessed 11-06-2024]},
}

@article{broholm2020quantum,
  title={Quantum spin liquids},
  author={Broholm, C and Cava, RJ and Kivelson, SA and Nocera, DG and Norman, MR and Senthil, T},
  journal={Science},
  volume={367},
  number={6475},
  pages={eaay0668},
  year={2020},
  publisher={American Association for the Advancement of Science}
}

@article{yin2022topological,
  title={Topological kagome magnets and superconductors},
  author={Yin, Jia-Xin and Lian, Biao and Hasan, M Zahid},
  journal={Nature},
  volume={612},
  number={7941},
  pages={647--657},
  year={2022},
  publisher={Nature Publishing Group UK London}
}

@article{slot2017experimental,
  title={Experimental realization and characterization of an electronic Lieb lattice},
  author={Slot, Marlou R and Gardenier, Thomas S and Jacobse, Peter H and Van Miert, Guido CP and Kempkes, Sander N and Zevenhuizen, Stephan JM and Smith, Cristiane Morais and Vanmaekelbergh, Daniel and Swart, Ingmar},
  journal={Nature physics},
  volume={13},
  number={7},
  pages={672--676},
  year={2017},
  publisher={Nature Publishing Group UK London}
}

@article{martinez1973archimedean,
  title={Archimedean lattices},
  author={Martinez, Jorge},
  journal={Algebra Universalis},
  volume={3},
  pages={247--260},
  year={1973},
  publisher={Springer}
}

@article{eddi2009archimedean,
  title={Archimedean lattices in the bound states of wave interacting particles},
  author={Eddi, A and Decelle, A and Fort, E and Couder, Y},
  journal={Europhysics Letters},
  volume={87},
  number={5},
  pages={56002},
  year={2009},
  publisher={IOP Publishing}
}

@article{kang2020topological,
  title={Topological flat bands in frustrated kagome lattice CoSn},
  author={Kang, Mingu and Fang, Shiang and Ye, Linda and Po, Hoi Chun and Denlinger, Jonathan and Jozwiak, Chris and Bostwick, Aaron and Rotenberg, Eli and Kaxiras, Efthimios and Checkelsky, Joseph G and others},
  journal={Nature communications},
  volume={11},
  number={1},
  pages={4004},
  year={2020},
  publisher={Nature Publishing Group UK London}
}

@article{tsai2015interaction,
  title={Interaction-driven topological and nematic phases on the Lieb lattice},
  author={Tsai, Wei-Feng and Fang, Chen and Yao, Hong and Hu, Jiangping},
  journal={New Journal of Physics},
  volume={17},
  number={5},
  pages={055016},
  year={2015},
  publisher={IOP Publishing}
}

@article{mukherjee2015observation,
  title={Observation of a localized flat-band state in a photonic Lieb lattice},
  author={Mukherjee, Sebabrata and Spracklen, Alexander and Choudhury, Debaditya and Goldman, Nathan and {\"O}hberg, Patrik and Andersson, Erika and Thomson, Robert R},
  journal={Physical review letters},
  volume={114},
  number={24},
  pages={245504},
  year={2015},
  publisher={APS}
}

@article{vicencio2015observation,
  title={Observation of localized states in Lieb photonic lattices},
  author={Vicencio, Rodrigo A and Cantillano, Camilo and Morales-Inostroza, Luis and Real, Basti{\'a}n and Mej{\'\i}a-Cort{\'e}s, Cristian and Weimann, Steffen and Szameit, Alexander and Molina, Mario I},
  journal={Physical review letters},
  volume={114},
  number={24},
  pages={245503},
  year={2015},
  publisher={APS}
}

@article{xie2018crystal,
  title={Crystal graph convolutional neural networks for an accurate and interpretable prediction of material properties},
  author={Xie, Tian and Grossman, Jeffrey C},
  journal={Physical review letters},
  volume={120},
  number={14},
  pages={145301},
  year={2018},
  publisher={APS}
}

@article{geiger2022e3nn,
  title={e3nn: Euclidean neural networks},
  author={Geiger, Mario and Smidt, Tess},
  journal={arXiv preprint arXiv:2207.09453},
  year={2022}
}

@article{chen2021direct,
  title={Direct prediction of phonon density of states with Euclidean neural networks},
  author={Chen, Zhantao and Andrejevic, Nina and Smidt, Tess and Ding, Zhiwei and Xu, Qian and Chi, Yen-Ting and Nguyen, Quynh T and Alatas, Ahmet and Kong, Jing and Li, Mingda},
  journal={Advanced Science},
  volume={8},
  number={12},
  pages={2004214},
  year={2021},
  publisher={Wiley Online Library}
}

@article{riebesell2023matbench,
  title={Matbench Discovery--An evaluation framework for machine learning crystal stability prediction},
  author={Riebesell, Janosh and Goodall, Rhys EA and Jain, Anubhav and Benner, Philipp and Persson, Kristin A and Lee, Alpha A},
  journal={arXiv preprint arXiv:2308.14920},
  year={2023}
}

@article{song2020score,
  title={Score-based generative modeling through stochastic differential equations},
  author={Song, Yang and Sohl-Dickstein, Jascha and Kingma, Diederik P and Kumar, Abhishek and Ermon, Stefano and Poole, Ben},
  journal={arXiv preprint arXiv:2011.13456},
  year={2020}
}

@inproceedings{lugmayr2022repaint,
  title={Repaint: Inpainting using denoising diffusion probabilistic models},
  author={Lugmayr, Andreas and Danelljan, Martin and Romero, Andres and Yu, Fisher and Timofte, Radu and Van Gool, Luc},
  booktitle={Proceedings of the IEEE/CVF conference on computer vision and pattern recognition},
  pages={11461--11471},
  year={2022}
}

@article{xie2021crystal,
  title={Crystal diffusion variational autoencoder for periodic material generation},
  author={Xie, Tian and Fu, Xiang and Ganea, Octavian-Eugen and Barzilay, Regina and Jaakkola, Tommi},
  journal={arXiv preprint arXiv:2110.06197},
  year={2021}
}

@article{jiao2024crystal,
  title={Crystal structure prediction by joint equivariant diffusion},
  author={Jiao, Rui and Huang, Wenbing and Lin, Peijia and Han, Jiaqi and Chen, Pin and Lu, Yutong and Liu, Yang},
  journal={Advances in Neural Information Processing Systems},
  volume={36},
  year={2024}
}

@article{jiao2024space,
  title={Space Group Constrained Crystal Generation},
  author={Jiao, Rui and Huang, Wenbing and Liu, Yu and Zhao, Deli and Liu, Yang},
  journal={arXiv preprint arXiv:2402.03992},
  year={2024}
}

@article{cao2024space,
  title={Space Group Informed Transformer for Crystalline Materials Generation},
  author={Cao, Zhendong and Luo, Xiaoshan and Lv, Jian and Wang, Lei},
  journal={arXiv preprint arXiv:2403.15734},
  year={2024}
}

@article{yang2023scalable,
  title={Scalable diffusion for materials generation},
  author={Yang, Mengjiao and Cho, KwangHwan and Merchant, Amil and Abbeel, Pieter and Schuurmans, Dale and Mordatch, Igor and Cubuk, Ekin Dogus},
  journal={arXiv preprint arXiv:2311.09235},
  year={2023}
}

@article{merchant2023scaling,
  title={Scaling deep learning for materials discovery},
  author={Merchant, Amil and Batzner, Simon and Schoenholz, Samuel S and Aykol, Muratahan and Cheon, Gowoon and Cubuk, Ekin Dogus},
  journal={Nature},
  volume={624},
  number={7990},
  pages={80--85},
  year={2023},
  publisher={Nature Publishing Group UK London}
}

@article{zeni2023mattergen,
  title={Mattergen: a generative model for inorganic materials design},
  author={Zeni, Claudio and Pinsler, Robert and Z{\"u}gner, Daniel and Fowler, Andrew and Horton, Matthew and Fu, Xiang and Shysheya, Sasha and Crabb{\'e}, Jonathan and Sun, Lixin and Smith, Jake and others},
  journal={arXiv preprint arXiv:2312.03687},
  year={2023}
}

@article{jain2013commentary,
  title={Commentary: The Materials Project: A materials genome approach to accelerating materials innovation},
  author={Jain, Anubhav and Ong, Shyue Ping and Hautier, Geoffroy and Chen, Wei and Richards, William Davidson and Dacek, Stephen and Cholia, Shreyas and Gunter, Dan and Skinner, David and Ceder, Gerbrand and others},
  journal={APL materials},
  volume={1},
  number={1},
  pages={011002},
  year={2013},
  publisher={American Institute of PhysicsAIP}
}

@article{zachariasen1973metallic,
  title={Metallic radii and electron configurations of the 5f- 6d metals},
  author={Zachariasen, WH},
  journal={Journal of Inorganic and Nuclear Chemistry},
  volume={35},
  number={10},
  pages={3487--3497},
  year={1973},
  publisher={Elsevier}
}

@article{ong2013python,
  title={Python Materials Genomics (pymatgen): A robust, open-source python library for materials analysis},
  author={Ong, Shyue Ping and Richards, William Davidson and Jain, Anubhav and Hautier, Geoffroy and Kocher, Michael and Cholia, Shreyas and Gunter, Dan and Chevrier, Vincent L and Persson, Kristin A and Ceder, Gerbrand},
  journal={Computational Materials Science},
  volume={68},
  pages={314--319},
  year={2013},
  publisher={Elsevier}
}

@article{davies2019smact,
  title={Smact: Semiconducting materials by analogy and chemical theory},
  author={Davies, Daniel W and Butler, Keith T and Jackson, Adam J and Skelton, Jonathan M and Morita, Kazuki and Walsh, Aron},
  journal={Journal of Open Source Software},
  volume={4},
  number={38},
  pages={1361},
  year={2019}
}

@article{pan2021benchmarking,
  title={Benchmarking coordination number prediction algorithms on inorganic crystal structures},
  author={Pan, Hillary and Ganose, Alex M and Horton, Matthew and Aykol, Muratahan and Persson, Kristin A and Zimmermann, Nils ER and Jain, Anubhav},
  journal={Inorganic chemistry},
  volume={60},
  number={3},
  pages={1590--1603},
  year={2021},
  publisher={ACS Publications}
}

@article{zimmermann2020local,
  title={Local structure order parameters and site fingerprints for quantification of coordination environment and crystal structure similarity},
  author={Zimmermann, Nils ER and Jain, Anubhav},
  journal={RSC advances},
  volume={10},
  number={10},
  pages={6063--6081},
  year={2020},
  publisher={Royal Society of Chemistry}
}

@article{kresse1996efficient,
  title={Efficient iterative schemes for ab initio total-energy calculations using a plane-wave basis set},
  author={Kresse, Georg and Furthm{\"u}ller, J{\"u}rgen},
  journal={Physical review B},
  volume={54},
  number={16},
  pages={11169},
  year={1996},
  publisher={APS}
}

@article{blochl1994projector,
  title={Projector augmented-wave method},
  author={Bl{\"o}chl, Peter E},
  journal={Physical review B},
  volume={50},
  number={24},
  pages={17953},
  year={1994},
  publisher={APS}
}

@article{kresse1999ultrasoft,
  title={From ultrasoft pseudopotentials to the projector augmented-wave method},
  author={Kresse, Georg and Joubert, Daniel},
  journal={Physical review b},
  volume={59},
  number={3},
  pages={1758},
  year={1999},
  publisher={APS}
}

@article{perdew1996generalized,
  title={Generalized gradient approximation made simple},
  author={Perdew, John P and Burke, Kieron and Ernzerhof, Matthias},
  journal={Physical review letters},
  volume={77},
  number={18},
  pages={3865},
  year={1996},
  publisher={APS}
}

@article{wang2021vaspkit,
  title={VASPKIT: A user-friendly interface facilitating high-throughput computing and analysis using VASP code},
  author={Wang, Vei and Xu, Nan and Liu, Jin-Cheng and Tang, Gang and Geng, Wen-Tong},
  journal={Computer Physics Communications},
  volume={267},
  pages={108033},
  year={2021},
  publisher={Elsevier}
}

@article{momma2011vesta,
  title={VESTA 3 for three-dimensional visualization of crystal, volumetric and morphology data},
  author={Momma, Koichi and Izumi, Fujio},
  journal={Journal of applied crystallography},
  volume={44},
  number={6},
  pages={1272--1276},
  year={2011},
  publisher={International Union of Crystallography}
}

@article{stukowski2009visualization,
  title={Visualization and analysis of atomistic simulation data with OVITO--the Open Visualization Tool},
  author={Stukowski, Alexander},
  journal={Modelling and simulation in materials science and engineering},
  volume={18},
  number={1},
  pages={015012},
  year={2009},
  publisher={IOP Publishing}
}

@Article{Hashimoto2014,
author={Hashimoto, Makoto
and Vishik, Inna M.
and He, Rui-Hua
and Devereaux, Thomas P.
and Shen, Zhi-Xun},
title={Energy gaps in high-transition-temperature cuprate superconductors},
journal={Nature Physics},
year={2014},
month={7},
day={01},
volume={10},
number={7},
pages={483-495},
issn={1745-2481},
doi={10.1038/nphys3009},
url={https://doi.org/10.1038/nphys3009}
}

@misc{figshare,
  title = {Structural Constraint Integration in Generative Model for Discovery of Quantum Material Candidates},
  author = {Okabe, Ryotaro},
  year = {2024},
  url = {https://doi.org/10.6084/m9.figshare.c.7283062.v1},
}

@Article{Checkelsky2024,
author={Checkelsky, Joseph G.
and Bernevig, B. Andrei
and Coleman, Piers
and Si, Qimiao
and Paschen, Silke},
title={Flat bands, strange metals and the Kondo effect},
journal={Nature Reviews Materials},
year={2024},
month={2},
day={20},
issn={2058-8437},
doi={10.1038/s41578-023-00644-z},
url={https://doi.org/10.1038/s41578-023-00644-z}
}

@Article{Speybroeck2015,
author ="Van Speybroeck, Veronique and Hemelsoet, Karen and Joos, Lennart and Waroquier, Michel and Bell, Robert G. and Catlow, C. Richard A.",
title  ="Advances in theory and their application within the field of zeolite chemistry",
journal  ="Chem. Soc. Rev.",
year  ="2015",
volume  ="44",
issue  ="20",
pages  ="7044-7111",
publisher  ="The Royal Society of Chemistry",
doi  ="10.1039/C5CS00029G",
url  ="http://dx.doi.org/10.1039/C5CS00029G"}

\end{document}


\newcommand{\beginsupplement}{%
        \setcounter{table}{0}
        \renewcommand{\thetable}{S\arabic{table}}%
        \setcounter{figure}{0}
        \renewcommand{\thefigure}{S\arabic{figure}}%
        \setcounter{equation}{0}
        \renewcommand{\theequation}{S\arabic{equation}}%
     }
\renewcommand{\hbar}{\mathchar'26\mkern-9mu h}
\beginsupplement
\newcounter{suppNoteCounter}
\setcounter{suppNoteCounter}{0}




























\maketitle
\newcommand{\beginsupplement}{%
        \setcounter{table}{0}
        \renewcommand{\thetable}{S\arabic{table}}%
        \setcounter{figure}{0}
        \renewcommand{\thefigure}{S\arabic{figure}}%
        \setcounter{equation}{0}
        \renewcommand{\theequation}{S\arabic{equation}}%
     }
\renewcommand{\hbar}{\mathchar'26\mkern-9mu h}
\beginsupplement
\tableofcontents
\newcounter{suppNoteCounter}
\setcounter{suppNoteCounter}{0}

\input{sections/si_text/si_data_structs/figs_tri}

\input{sections/si_text/si_data_structs/figs_hon}



\input{sections/si_text/si_data_structs/figs_elt}

\input{sections/si_text/si_data_structs/figs_sns}

\input{sections/si_text/si_data_structs/figs_tsq}

\input{sections/si_text/si_data_structs/figs_srt}

\input{sections/si_text/si_data_structs/figs_snh}

\input{sections/si_text/si_data_structs/figs_trh}

\input{sections/si_text/si_data_structs/figs_grt}

\input{sections/si_text/si_data_structs/figs_lieb}


\nocitesupp{*}
\bibliographystylesupp{abbrv}
\bibliographysupp{si}


\maketitle
\newcommand{\beginsupplement}{%
        \setcounter{table}{0}
        \renewcommand{\thetable}{S\arabic{table}}%
        \setcounter{figure}{0}
        \renewcommand{\thefigure}{S\arabic{figure}}%
        \setcounter{equation}{0}
        \renewcommand{\theequation}{S\arabic{equation}}%
     }
\renewcommand{\hbar}{\mathchar'26\mkern-9mu h}
\beginsupplement
\tableofcontents
\newcounter{suppNoteCounter}
\setcounter{suppNoteCounter}{0}

\newpage

\stepcounter{suppNoteCounter}
\section{Archimedean and Lieb lattices as geometrical pattern constraints}

Archimedean lattices (ALs)\cite{martinez1973archimedean, eddi2009archimedean}, commonly referred to as Archimedean tilings, are distinctive for their planar, uniform tiling, where each vertex configuration is identical. Unlike regular tilings, which utilize only one type of regular polygon, Archimedean lattices incorporate multiple types of regular polygons but are arranged uniformly at each vertex. A key feature of ALs is their vertex-transitivity, which allows any vertex to be mapped to any other through a series of reflections, rotations, and translations, thus preserving the arrangement's overall symmetry.

There are exactly 11 types of ALs, each uniquely defined by the types and sequences of polygons that meet at each vertex. Each AL can be described both as a descriptive name and the numerical name called by the list of the polygons surrounding one vertex. These include the Triangular ($3^6$), Honeycomb ($6^3$), and Kagome ($3,6,3,6$) lattices, which are composed of triangles and hexagons in different configurations. The Square lattice ($4^4$) consists solely of squares. The Elongated triangular ($3^3,4^2$) and Snub square ($3^2,4,3,4$) lattices mix triangles and squares in varied layouts. Other forms include the Truncated square ($4,8^2$), Small rhombitrihexagonal ($3,4,6,4$), Snub hexagonal ($3^4,6$), Truncated hexagonal ($3,12^2$), and the Great rhombitrihexagonal ($4,6,12$), which involve combinations of squares, hexagons, and dodecagons, each offering complex geometric arrangements. 

The Lieb lattice is a unique two-dimensional lattice structure characterized by its three sites per unit cell configured in a square shape. The vertices of this lattice are positioned at each corner and at the midpoint of each edge, forming a bipartite lattice. This specific arrangement allows the lattice to be divided into two interpenetrating sublattices, where each site on one only interacts with sites on the other sublattice. The configuration of the Lieb lattice is particularly significant in research areas focusing on optical, magnetic, and transport properties due to its potential for facilitating unusual localized states. These attributes make the Lieb lattice a valuable model in theoretical physics and the practical development of materials with tailored electronic properties.

\begin{figure}[H]
\includegraphics[width=0.8\textwidth]{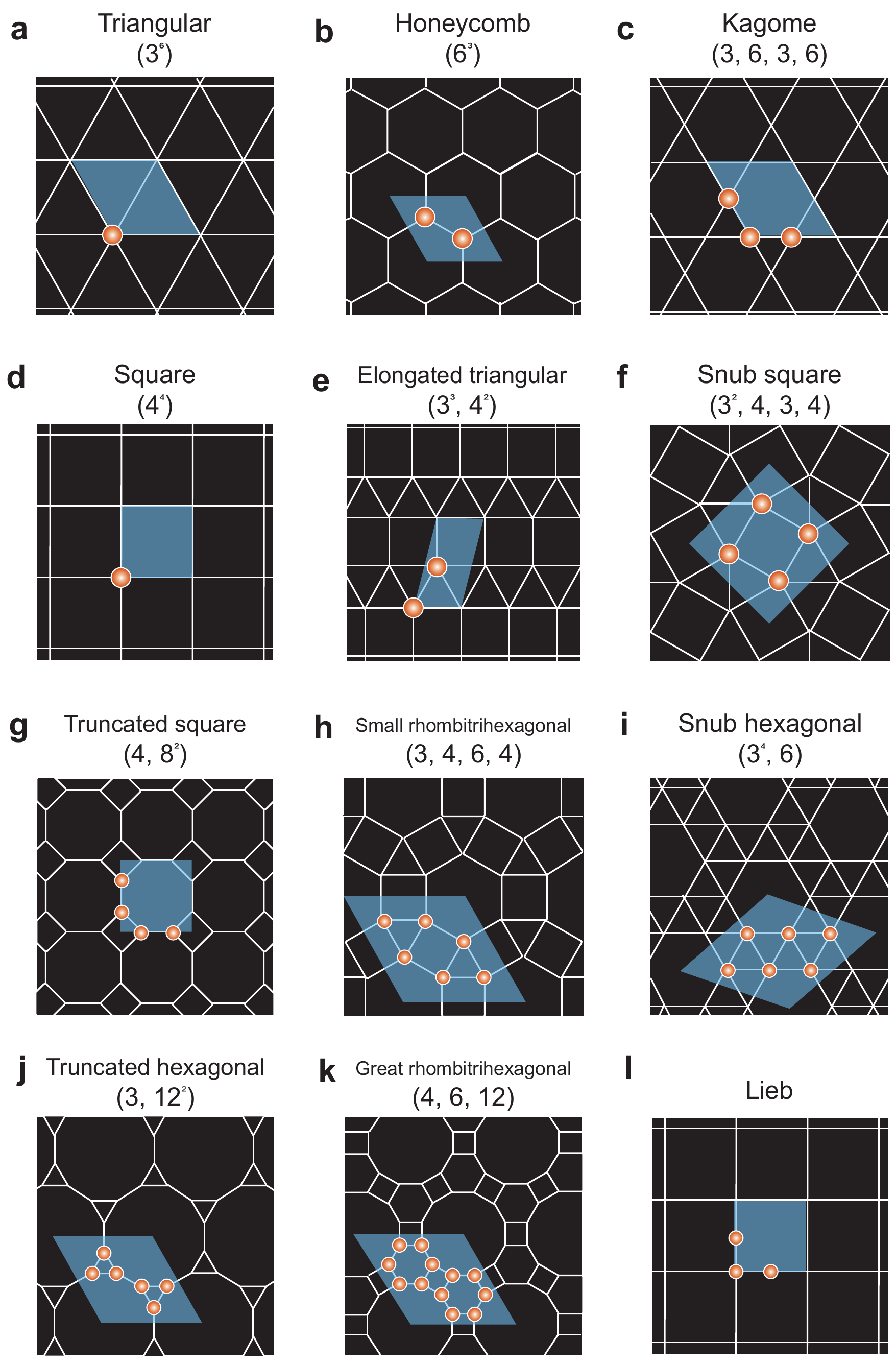}
\centering
\caption{
\textbf{The geometric patterns of Archimedean and Lieb lattices.}  
The geometric pattern is drawn with white lines. The unit cells (blue shaded area) contain nodes (orange dots) for each lattice type. \textbf{a.} Triangular  \textbf{b.} Honeycomb \textbf{c.} Kagome \textbf{d.} Square \textbf{e.} Elongated triangular \textbf{f.} Snub square \textbf{g.} Truncated square \textbf{h.} Small rhombitrihexagonal \textbf{i.} Snub hexagonal \textbf{j.} Truncated hexagonal \textbf{k.} Great rhombitrihexagonal \textbf{l.} Lieb lattice. 
}
\label{SI_arch_lat}
\end{figure}

Table \ref{tab_arch_latt} lists the characteristics of Archimedean and Lieb-like lattices. It covers the header that we used for giving file names to each output material, the property of the unit cell, the number of nodes forming AL per unit cell $N^c$, and the fractional coordinates. Here, we write six lattice parameters as $l_1$, $l_2$, $l_3$, $\alpha$, $\beta$, $\gamma$. SCIGEN imposes constraints for the lattice parameters that reflect the AL structures ($l_1$, $l_2$, $\gamma$), while there are no constraints on the other lattice parameters ($l_3$, $\alpha$, $\beta$). We also included the constants $k_{latt} = l_1/d^c$ (or $l_2/d^c$) as the ratio of lattice vector length ($l_1$, $l_2$) to the bond length between neighboring vertices ($d^c$), indicating the relative size of the AL unit cell. Figure \ref{SI_arch_lat} visualizes all types of AL and a Lieb-like lattice. The unit cell area is highlighted with blue areas, and the red vertices represent the required positions within the unit cell. \\

\begin{table}[H]
\begin{center}
\caption{\textbf{Profile of Archimedean lattice and Lieb Lattice}}
\label{tab_arch_latt}
\begin{tabular}{c*{5}{c}m{5cm}}
            Archimedean lattice &  header & Unit cell & $N^{c}$ & $k_{latt}$ & $F_0^{c}$  \\
\hline
\hline
\begin{tabular}[t]{@{}c@{}} Triangular\\$(3^6)$\end{tabular} & tri &\begin{tabular}[t]{@{}c@{}} Hexagonal\\ ($l_1=l_2$, $\gamma=2\pi/3$)\end{tabular} & 1  & 1.0000 & (0.0000, 0.0000, z)  \\
\hline
\begin{tabular}[t]{@{}c@{}}Honeycomb\\$(6^3)$\end{tabular} & hon & \begin{tabular}[t]{@{}c@{}} Hexagonal\\ ($l_1=l_2$, $\gamma=2\pi/3$)\end{tabular} & 2 & 1.7321 & \begin{tabular}[t]{@{}c@{}}(0.3333, 0.6667, z)\\ (0.6667, 0.3333, z)\end{tabular}  \\  
\hline
\begin{tabular}[t]{@{}c@{}}Kagome\\$(3, 6, 3, 6)$\end{tabular} & kag & \begin{tabular}[t]{@{}c@{}} Hexagonal\\ ($l_1=l_2$, $\gamma=2\pi/3$)\end{tabular} & 3 & 2.0000 & \begin{tabular}[t]{@{}c@{}}(0.0000, 0.0000, z)\\ (0.5000, 0.0000, z)\\ (0.0000, 0.5000, z)\end{tabular}  \\
\hline
\begin{tabular}[t]{@{}c@{}}Square\\$(4^4)$\end{tabular} & sqr & \begin{tabular}[t]{@{}c@{}}Square\\ ($l_1=l_2$, $\gamma=\pi/2$)\end{tabular} & 1 & 1.0000 & (0.0000, 0.0000, z)  \\
\hline
\begin{tabular}[t]{@{}c@{}}Elongated triangular\\$(3^3, 4^2)$\end{tabular} & elt & \begin{tabular}[t]{@{}c@{}}Parallelogram\\ ($l_2=1.932l_1$, $\gamma=\pi/3$)\end{tabular} & 2 & \begin{tabular}[t]{@{}c@{}}1.0000, \\ 1.9319\end{tabular} & \begin{tabular}[t]{@{}c@{}}(0.0000, 0.0000, z)\\ (0.2679, 0.4641, z)\end{tabular}  \\
\hline
\begin{tabular}[t]{@{}c@{}} Snub square\\$(3^2, 4, 3, 4)$ \end{tabular} & sns & \begin{tabular}[t]{@{}c@{}}Square\\ ($l_1=l_2$, $\gamma=\pi/2$)\end{tabular} & 4 & 1.9319 & \begin{tabular}[t]{@{}c@{}}(0.3170, 0.1830, z)\\ (0.1830, 0.6830, z)\\ (0.6830, 0.8170, z)\\ (0.8170, 0.3170, z)\end{tabular}  \\
\hline
\begin{tabular}[t]{@{}c@{}} Truncated square\\$(4, 8^2)$ \end{tabular}  & tsq & \begin{tabular}[t]{@{}c@{}}Square\\ ($l_1=l_2$, $\gamma=\pi/2$)\end{tabular} & 4 & 2.4142 & \begin{tabular}[t]{@{}c@{}}(0.2929, 0.0000, z)\\ (0.7071, 0.0000, z)\\ (0.0000, 0.2929, z)\\ (0.0000, 0.7071, z)\end{tabular}  \\
\hline
\begin{tabular}[t]{@{}c@{}} Small rhombitrihexagonal\\$(3, 4, 6, 4)$ \end{tabular} & srh & \begin{tabular}[t]{@{}c@{}} Hexagonal\\ ($l_1=l_2$, $\gamma=2\pi/3$)\end{tabular} & 6 & 2.7321 & \begin{tabular}[t]{@{}c@{}}(0.4226, 0.2113, z)\\ (0.7887, 0.2113, z)\\ (0.7887, 0.5774, z)\\ (0.2113, 0.4226, z)\\ (0.2113, 0.7887, z)\\ (0.5774, 0.7887, z)\end{tabular}  \\
\hline
\begin{tabular}[t]{@{}c@{}} Snub hexagonal\\$(3^4, 6)$ \end{tabular} & snh & \begin{tabular}[t]{@{}c@{}} Hexagonal\\ ($l_1=l_2$, $\gamma=2\pi/3$)\end{tabular} & 6 & 2.6458 & \begin{tabular}[t]{@{}c@{}}(0.4286, 0.1429, z)\\ (0.8571, 0.2857, z)\\ (0.2857, 0.4286, z)\\ (0.7143, 0.5714, z)\\ (0.1429, 0.7143, z)\\ (0.5714, 0.8571, z)\end{tabular}  \\
\hline
\begin{tabular}[t]{@{}c@{}} Truncated hexagonal\\$(3, 12^2)$ \end{tabular} & trh & \begin{tabular}[t]{@{}c@{}} Hexagonal\\ ($l_1=l_2$, $\gamma=2\pi/3$)\end{tabular} & 6 & 3.7321 & \begin{tabular}[t]{@{}c@{}}(0.5774, 0.1547, z)\\ (0.8453, 0.4226, z)\\ (0.5774, 0.4226, z)\\ (0.1547, 0.5774, z)\\ (0.4226, 0.8453, z)\\ (0.4226, 0.5774, z)\end{tabular}  \\
\hline
\begin{tabular}[t]{@{}c@{}} Great rhombitrihexagonal\\$(4, 6, 12)$ \end{tabular} & grt  & \begin{tabular}[t]{@{}c@{}} Hexagonal\\ ($l_1=l_2$, $\gamma=2\pi/3$)\end{tabular} & 12 & 4.7321 & \begin{tabular}[t]{@{}c@{}}(0.8780, 0.3333, z)\\ (0.8780, 0.5447, z)\\ (0.6667, 0.5447, z)\\ (0.4553, 0.3333, z)\\ (0.4553, 0.1220, z)\\ (0.6667, 0.1220, z) \\ (0.3333, 0.8780, z)\\ (0.5447, 0.8780, z)\\ (0.5447, 0.6667, z)\\ (0.3333, 0.4553, z)\\ (0.1220, 0.4553, z)\\ (0.1220, 0.6667, z)\end{tabular}  \\
\hline
\begin{tabular}[t]{@{}c@{}} Lieb \end{tabular} & lieb  & \begin{tabular}[t]{@{}c@{}}Square\\ ($l_1=l_2$, $\gamma=\pi/2$)\end{tabular} & 3 & 2.0000 & \begin{tabular}[t]{@{}c@{}}(0.0000, 0.0000, z)\\ (0.5000, 0.0000, z)\\ (0.0000, 0.5000, z)\end{tabular}  \\

\hline
\end{tabular}
\end{center}
\end{table}

\stepcounter{suppNoteCounter}
\section{Initialization of the constraint of structures}

\begin{figure}[H]
\includegraphics[width=0.75\textwidth]{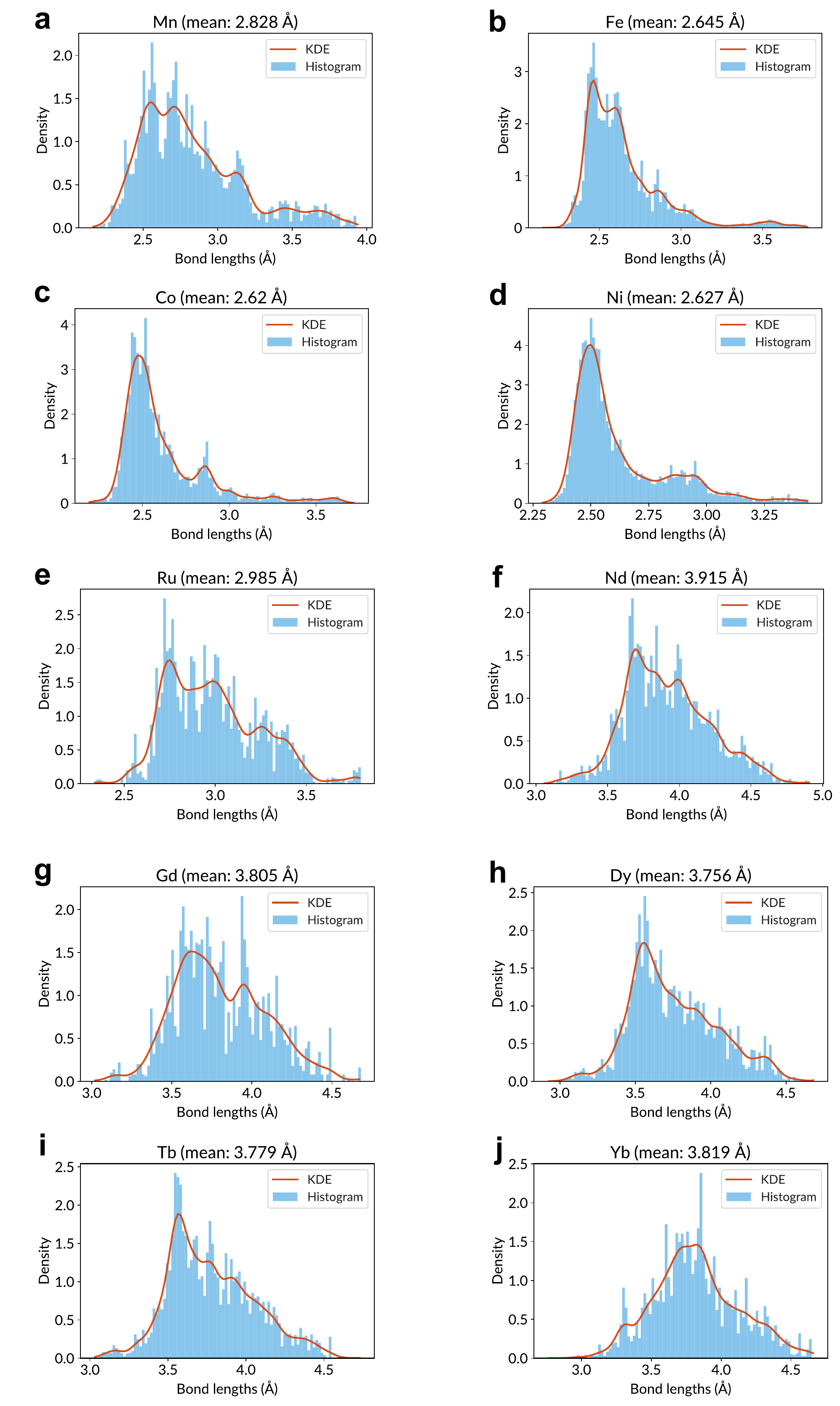}
\centering
\caption{
\textbf{The distribution of bond length $p_{d^c}(\mathcal{A}^c)$ sampled from MP-20 dataset.} 
The same atom types located in nearest-neighbor positions are captured, and their distances are measured. The blue histogram (blue) and the red line indicate the sampled distribution and the KDE-fitted curve for \textbf{a.} Mn, \textbf{b.} Fe, \textbf{c.} Co, \textbf{d.} Ni, \textbf{e.} Ru, \textbf{f.} Nd, \textbf{g.} Gd, \textbf{h.} Tb, \textbf{i.} Dy, \textbf{j.} Yb.
}

\label{SI_bond_len}
\end{figure}

We seek to give constraints to the diffusion model so that the output materials have Archimedean lattice (AL) structures formed with magnetic atoms. Here we present the detailed method of initializing the AL layers, which supplements Fig. 1b of the main article. To begin with, we can arbitrarily choose the AL type we want to impose. In our approach to discovering new materials with unique magnetic characteristics, we aim to place magnetic atoms at the vertices of the AL structures. Specifically, we select ten candidate atoms for these vertices: Mn, Fe, Co, Ni, Ru, Nd, Gd, Tb, Dy, and Yb. Once the vertex atom types $\mathcal{A}^c$ are sampled, their known magnetic properties and electronic configuration affect the coordination environment, which guides the generative process to form stable lattice structures. Next, we consider the bond lengths $d^c$ of vertices within the lattice, which highly depend on the atom type $\mathcal{A}^c$ of the AL vertices. Using CrystalNN\cite{pan2021benchmarking}, we identify the bonds of nearest neighbors and then employ kernel density estimation (KDE) to fit the sampled bond lengths, taking into account the metallic radii\cite{zachariasen1973metallic} of the atoms as the minimum thresholds. Figure \ref{SI_bond_len} illustrates the sampled distribution $p_{d^c}(\mathcal{A}^c)$ of nearest-neighbor distances $d^c$ for each magnetic atom type $\mathcal{A}^c$, providing a comprehensive overview of the expected bond lengths. The number of atoms per unit cell, denoted as $N$, is another crucial parameter. We generate a profile of $N$ by sampling from a normal distribution. For each AL and magnetic atom type, we sampled 3000 materials from a uniform distribution with $N$ ranging from $N^{c}+1$ to $N_{\text{max}}$. Here, $N^{c}$ is the number of vertices forming the unit cell of AL lattices. $N_{\text{max}}$ is the maximum number of atoms per unit cell, and we set it to 20 for profiling the $N$ distribution. 

\begin{figure}[H]
\includegraphics[width=0.8\textwidth]{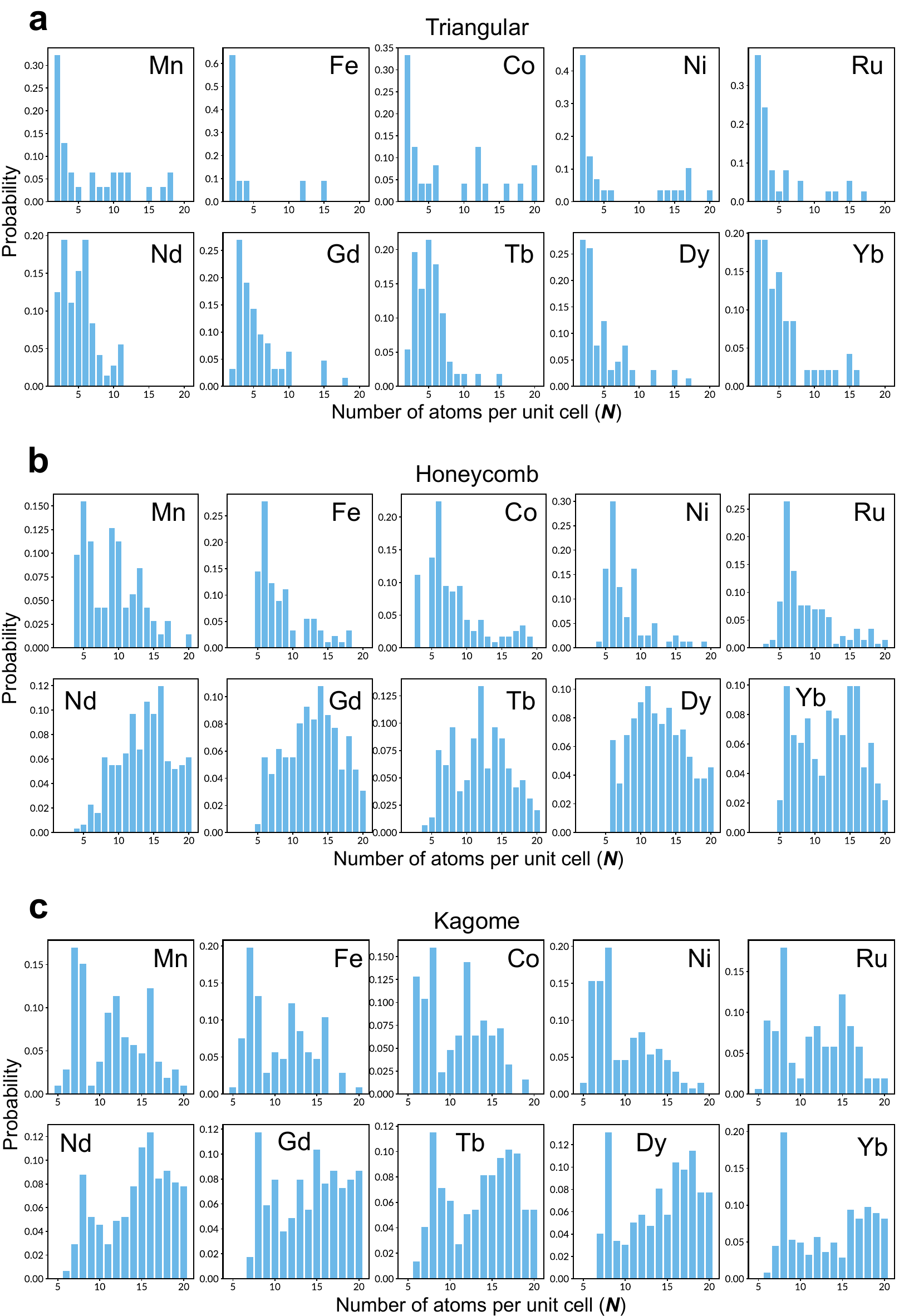}
\centering
\caption{
\textbf{The probability distribution $p_N(\mathcal{A}^c)$ for the number of atoms per unit cell $N$ in case of $\mathcal{A}^c$ as AL vertices.}  \\
3000 materials are generated for each type of lattice and each type of magnetic atoms. $N \in[N^c+1, 20]$ values are sampled from the uniform distribution, and pre-screening filters are applied to evaluate stable materials. \textbf{a.} Triangular  \textbf{b.} Honeycomb \textbf{c.} Kagome.
}

\label{SI_natm_partial_1}
\end{figure}

\begin{figure}[H]
\includegraphics[width=0.8\textwidth]{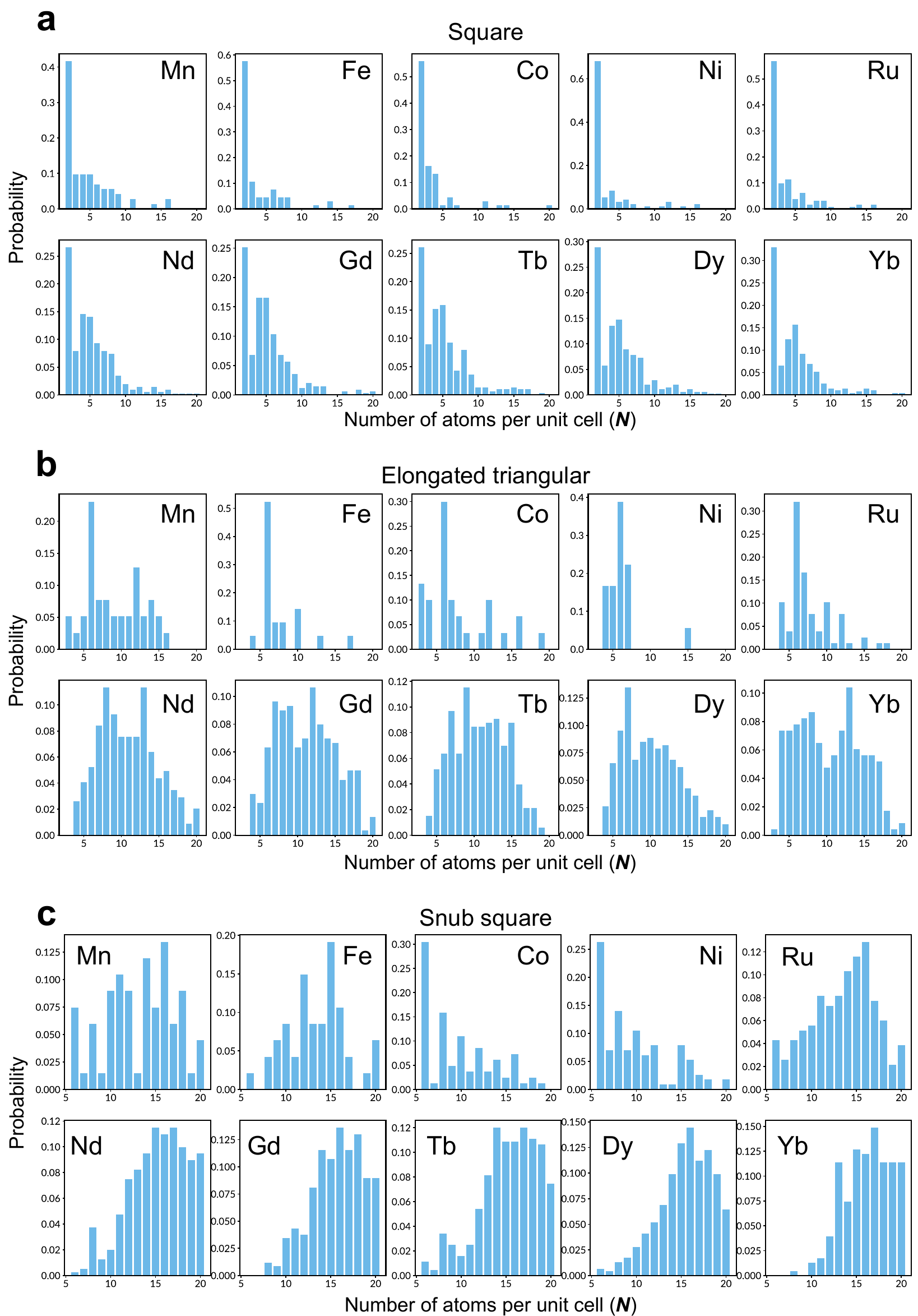}
\centering
\caption{
\textbf{The probability distribution $p_N(\mathcal{A}^c)$ for the number of atoms per unit cell $N$ in case of $\mathcal{A}^c$ as AL vertices.}  \\
3000 materials are generated for each type of lattice and each type of magnetic atoms. $N \in[N^c+1, 20]$ values are sampled from the uniform distribution, and pre-screening filters are applied to evaluate stable materials. \textbf{a.} Square \textbf{b.} Elongated triangular \textbf{c.} Snub square.
}

\label{SI_natm_partial_2}
\end{figure}

\begin{figure}[H]
\includegraphics[width=0.8\textwidth]{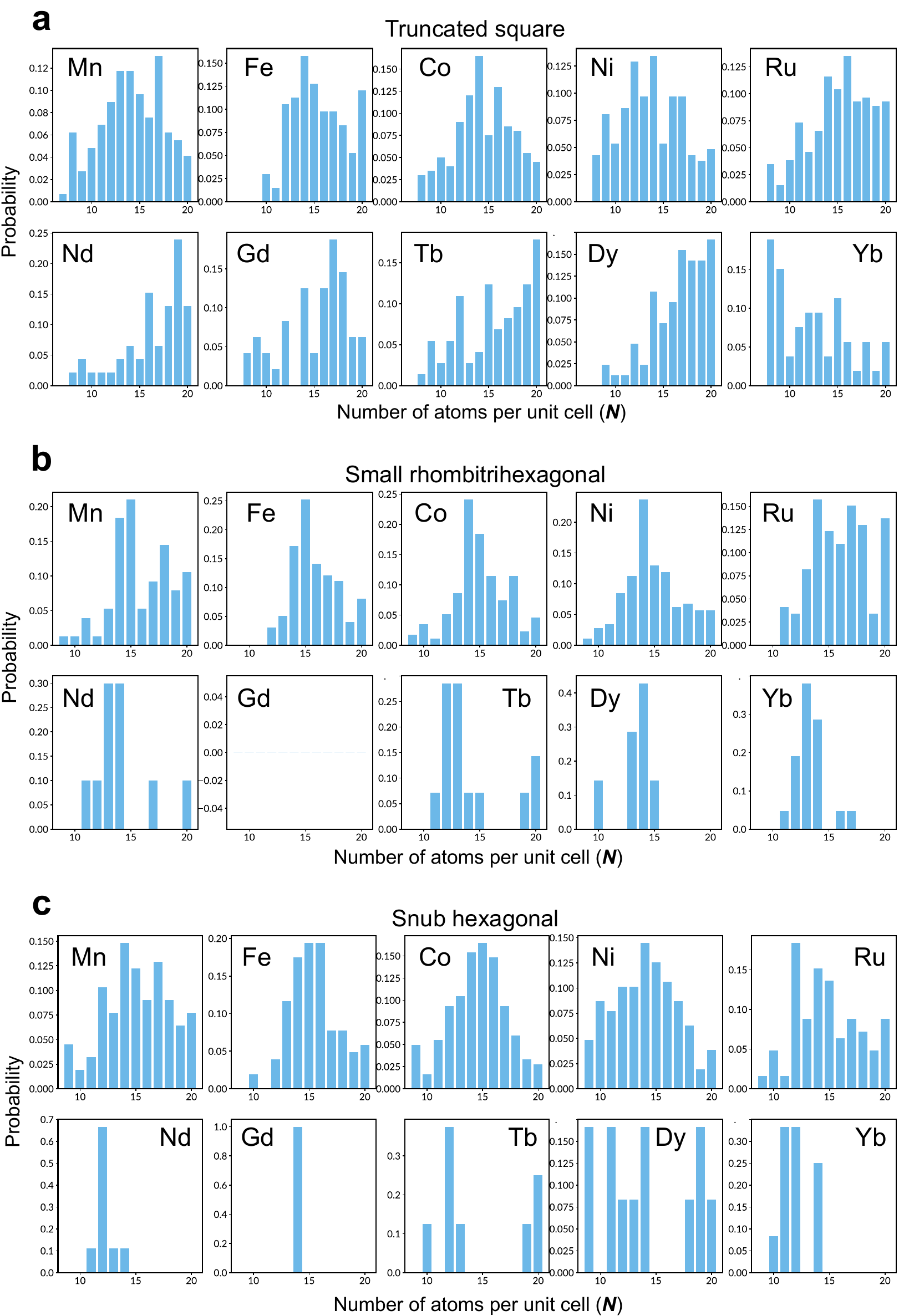}
\centering
\caption{
\textbf{The probability distribution $p_N(\mathcal{A}^c)$ for the number of atoms per unit cell $N$ in case of $\mathcal{A}^c$ as AL vertices.}  \\
3000 materials are generated for each type of lattice and each type of magnetic atoms. $N \in[N^c+1, 20]$ values are sampled from the uniform distribution, and pre-screening filters are applied to evaluate stable materials. \textbf{a.} Truncated square \textbf{b.} Small rhombitrihexagonal \textbf{c.} Snub hexagonal. 
}

\label{SI_natm_partial_3}
\end{figure}

\begin{figure}[H]
\includegraphics[width=0.8\textwidth]{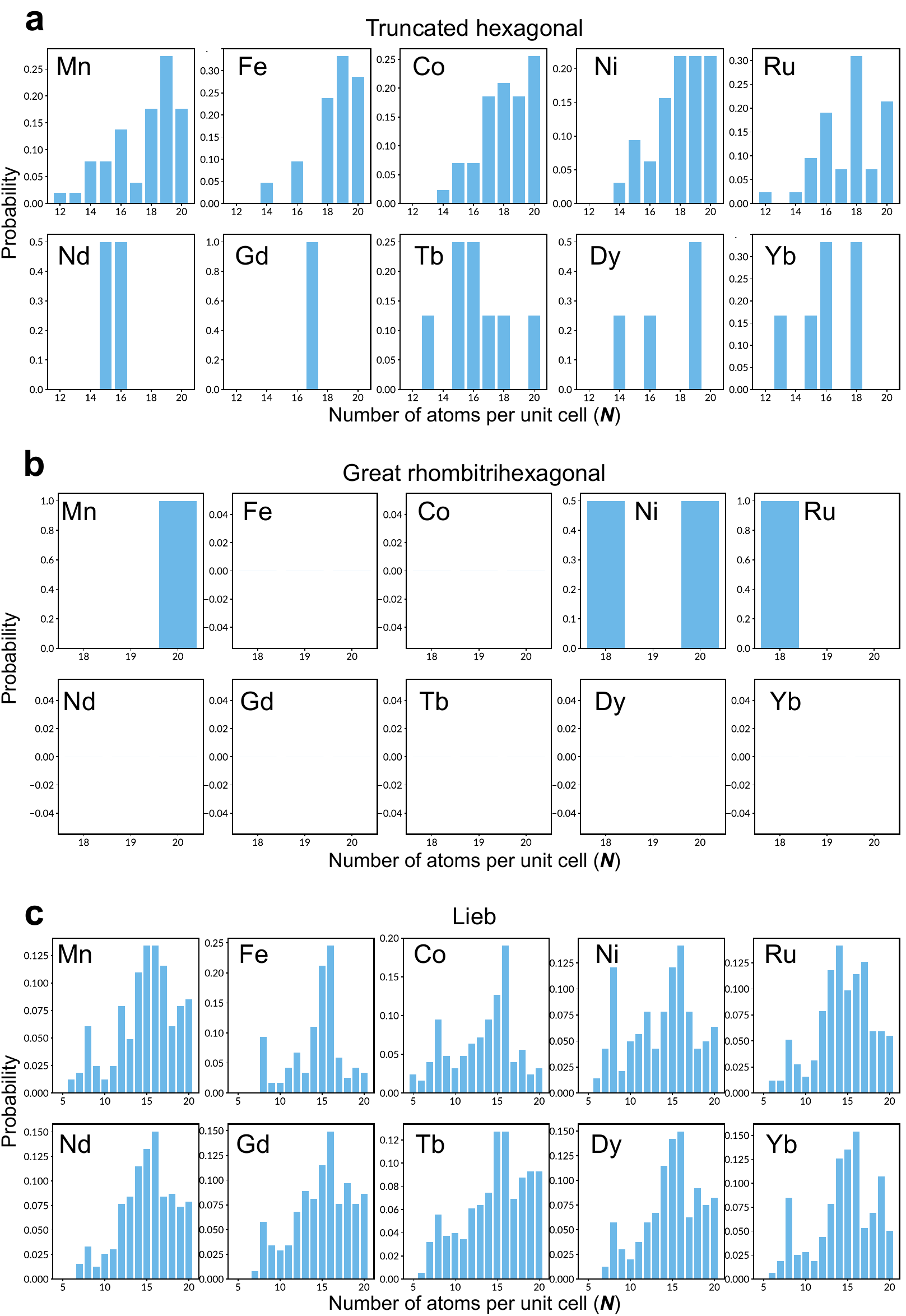}
\centering
\caption{
\textbf{The probability distribution $p_N(\mathcal{A}^c)$ for the number of atoms per unit cell $N$ in case of $\mathcal{A}^c$ as AL vertices.}  \\
3000 materials are generated for each type of lattice and each type of magnetic atoms. $N \in[N^c+1, 20]$ values are sampled from the uniform distribution, and pre-screening filters are applied to evaluate stable materials. \textbf{a.} Truncated hexagonal \textbf{b.} Great rhombitrihexagonal \textbf{c.} Lieb.
}

\label{SI_natm_partial_4}
\end{figure}

\begin{figure}[H]
\includegraphics[width=0.8\textwidth]{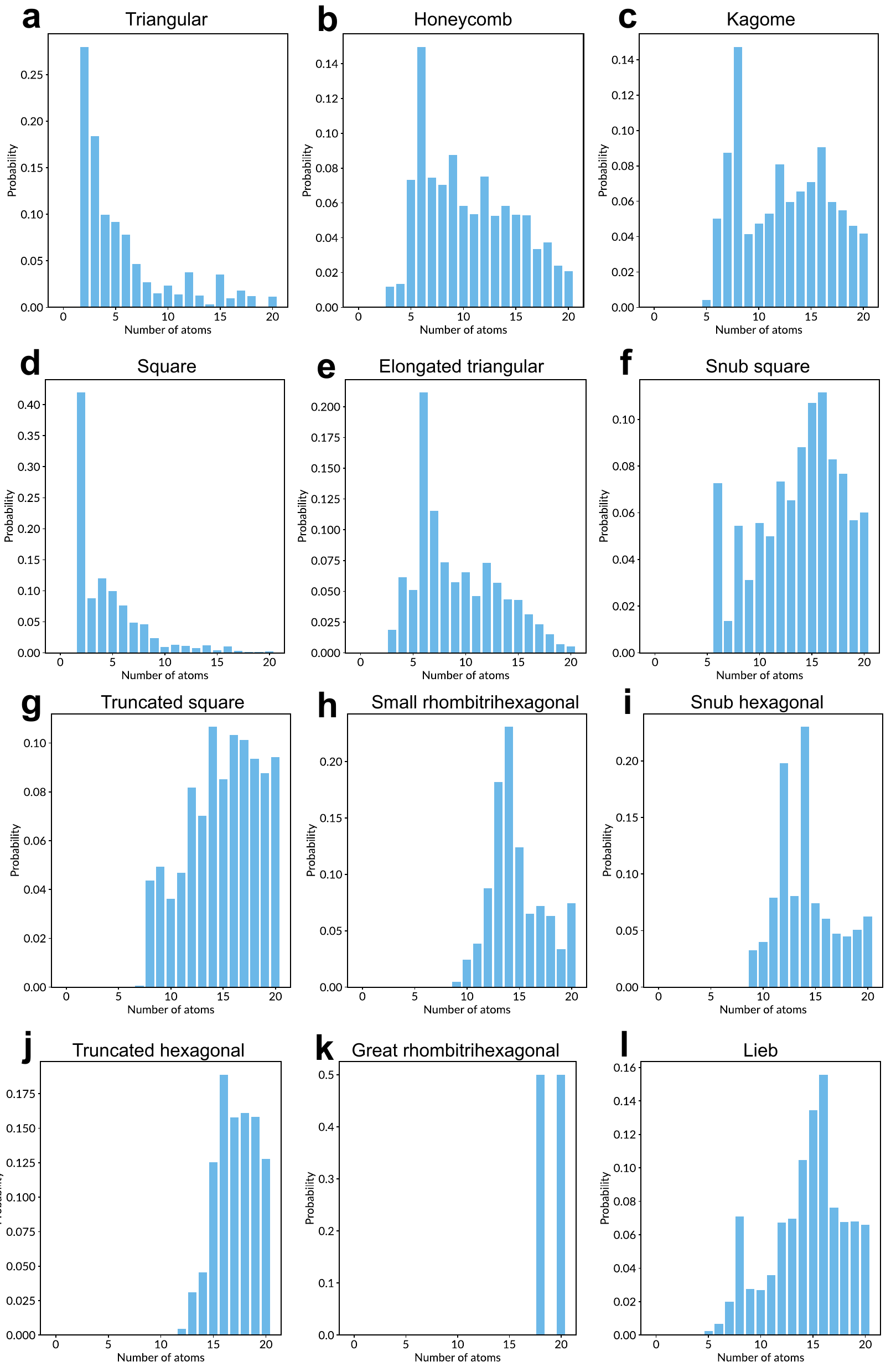}
\centering
\caption{
\textbf{The probability distribution $p_N$ of the number of atoms per unit cell $N$.}     
The probability distributions $p_N(\mathcal{A}^c)$ are averaged over all magnetic atom types $\mathcal{A}^c$. \textbf{a.} Triangular  \textbf{b.} Honeycomb \textbf{c.} Kagome \textbf{d.} Square \textbf{e.} Elongated triangular \textbf{f.} Snub square \textbf{g.} Truncated square \textbf{h.} Small rhombitrihexagonal \textbf{i.} Snub hexagonal \textbf{j.} Truncated hexagonal \textbf{k.} Great rhombitrihexagonal \textbf{l.} Lieb.
}

\label{SI_natm_count_total}
\end{figure}

\begin{figure}[H]
\includegraphics[width=0.8\textwidth]{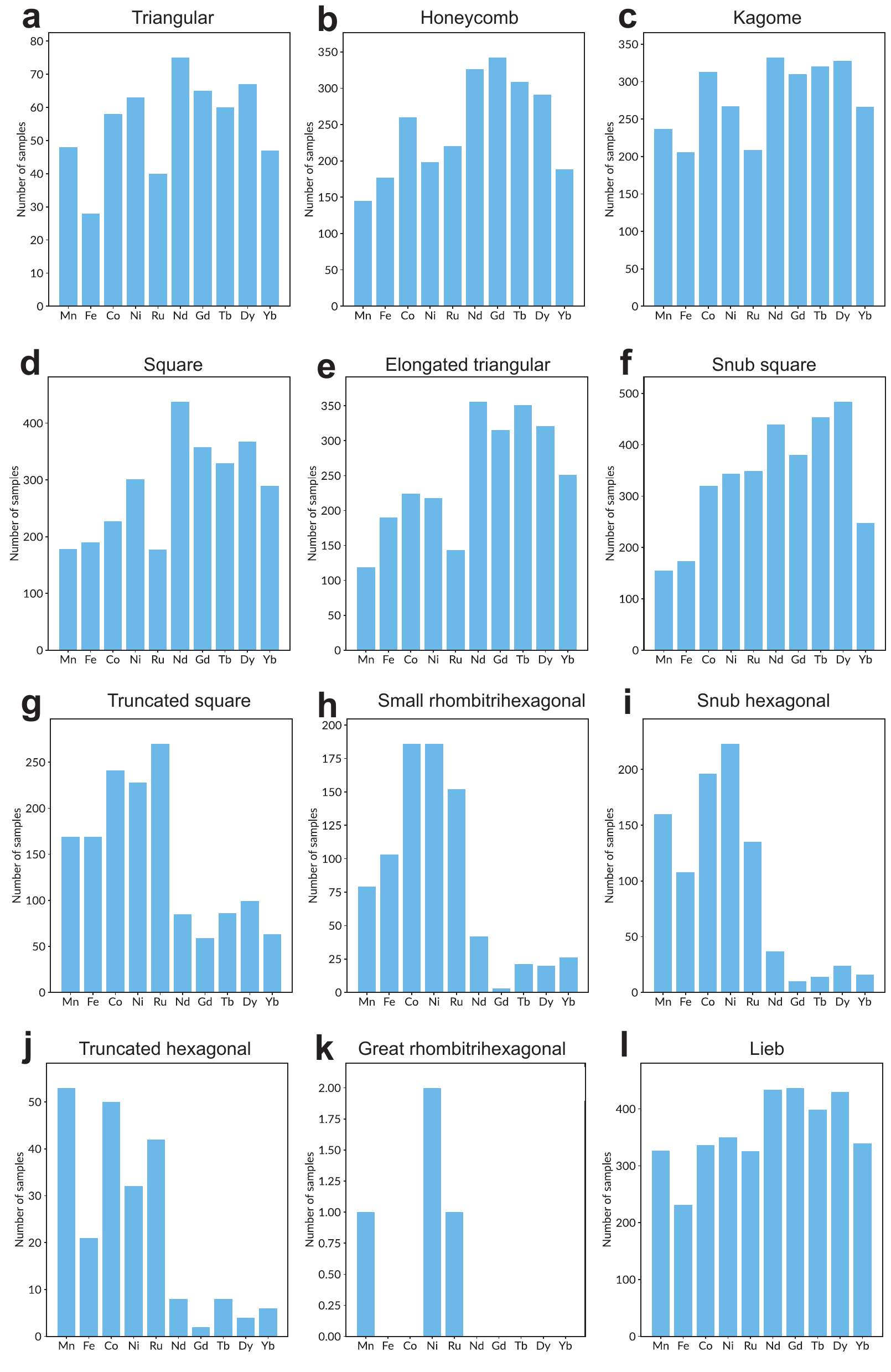}
\centering
\caption{
\textbf{Atom type distribution.}     
The survival ratio was calculated for each of the magnetic atoms forming AL structures. \textbf{a.} Triangular  \textbf{b.} Honeycomb \textbf{c.} Kagome \textbf{d.} Square \textbf{e.} Elongated triangular \textbf{f.} Snub square \textbf{g.} Truncated square \textbf{h.} Small rhombitrihexagonal \textbf{i.} Snub hexagonal \textbf{j.} Truncated hexagonal \textbf{k.} Great rhombitrihexagonal \textbf{l.} Lieb
}

\label{SI_atype_count_total}
\end{figure}

To ensure the generated materials were stable, we apply pre-screening filters to evaluate the reasonableness of the chosen parameters. These filters measure the survival ratios based on charge neutrality, space occupation ratio, and GNN-based classification models. Detailed explanations of the pre-screening procedures are provided in the Methods section and Supplementary Information 5. Figures \ref{SI_natm_partial_1}-\ref{SI_natm_partial_4} present $p_N(\mathcal{A}^c)$, the probability distribution of $N$ presenting stable materials with magnetic atom $\mathcal{A}^c$ as AL vertices. Note that $p_N(\mathcal{A}^c)$ is normalized, so that $\sum_N p_N(\mathcal{A}^c) =1$. By averaging the histogram $p_N(\mathcal{A}^c)$ over all $\mathcal{A}^c$ cases, we build a probability distribution of $N$ as $p_N = \text{mean}_{\mathcal{A}^c}(p_N(\mathcal{A}^c))$. We can sample $N$ from probability distribution $p_N$, which equally takes all of magnetic atom types $\mathcal{A}^c$ into account. Figure \ref{SI_natm_count_total} shows $p_N$ as the sampling profile of $N$ for initialization of the generation process. Furthermore, Figure \ref{SI_atype_count_total} presents the survival ratio of materials for each magnetic atom type as vertices. By rigorously specifying the AL types and sampling the vertex atom types from a uniform distribution, we generate structures with magnetic atoms forming AL structures.

\stepcounter{suppNoteCounter}
\section{The details of materials generation with geometrical constraint}

\begin{figure}[H]
\includegraphics[width=0.8\textwidth]{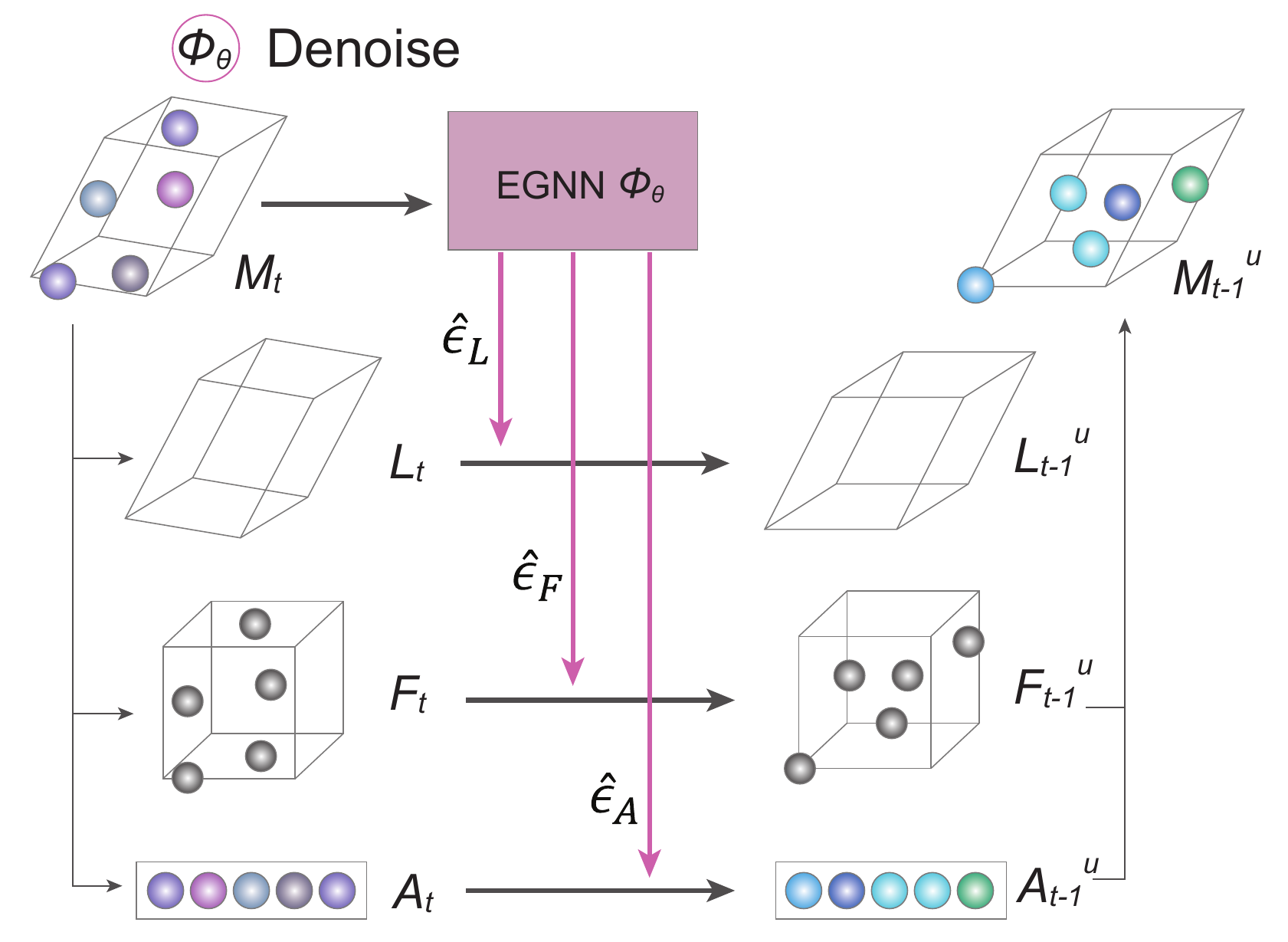}
\centering
\caption{
\textbf{Method of generative model for denoising material.} 
We apply the denoising term prediction implemented in DiffCSP. We input a diffused crystal structure $\bs{M}_t^u$ to get the denoising term for the lattice, the fractional coordinates, and the atom type. 
}

\label{SI_model_denoise}
\end{figure}

Figure 1 in the main manuscript presents the detailed algorithm of our material generation method with structural constraints. Our generative model is implemented based on DiffCSP\cite{jiao2024crystal}. Figure \ref{SI_model_denoise} elucidates the overview of the denoising method used in this process. Initially, we input the structure $\bs{M}_t^u=(\bs{L}_t, \bs{F}_t, \bs{A}_t)$, where each component is denoised to become $\bs{M}_{t-1}^u=(\bs{L}_{t-1}^u, \bs{F}_{t-1}^u, \bs{A}_{t-1}^u)$. This transformation is achieved by predicting the denoising terms $\hat{\bs{\epsilon}}_{\bs{F}}$, $\hat{\bs{\epsilon}}_{\bs{L}}$, and $\hat{\bs{\epsilon}}_{\bs{A}}$.

\begin{figure}[H]
\includegraphics[width=0.8\textwidth]{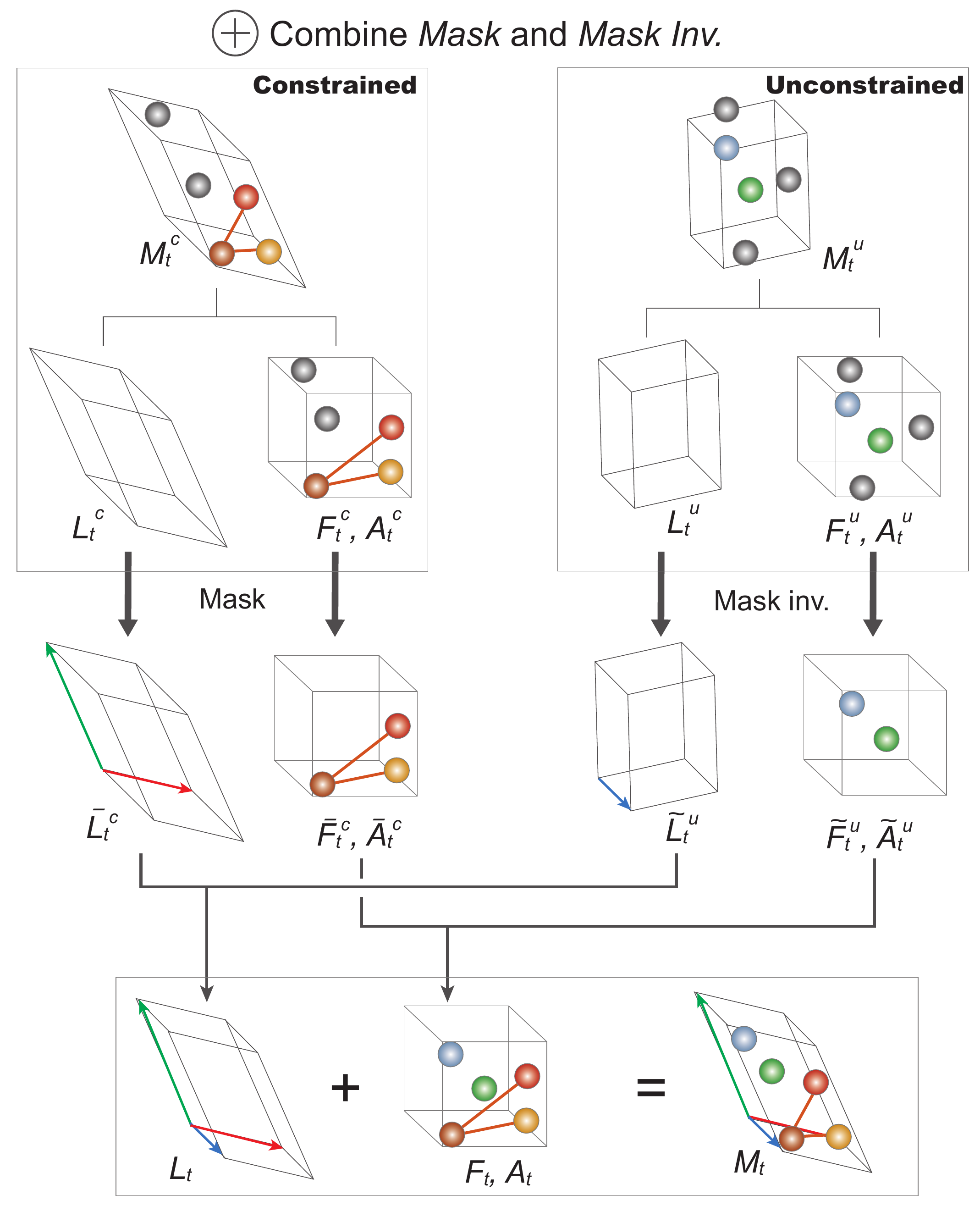}
\centering
\caption{
\textbf{Method of generative model for denoising material.} 
Both the constrained structures $\bs{M}_{t}^c$ and unconstrained ones $\bs{M}_{t}^u$ are decomposed to the lattice, atom types, and coordinates. After applying the masks to select the components to care for, structural components are integrated. 
}

\label{SI_model_concat}
\end{figure}

Figure \ref{SI_model_concat} illustrates the procedure to integrate the constrained structure with the unconstrained components to guide the material generation forming AL structures. For each timestamp $t$ of the denoising process, whenever we have $\bs{M}_{t}^u$, we obtain $\bs{M}_{t}$ by guiding the structure with the constrained component $\bs{M}_{t}^c$. The constrained structure is represented as $\bs{M}_t^c=(\bs{L}_t^c, \bs{F}_t^c, \bs{A}_t^c)$, and the unconstrained structure is $\bs{M}_t^u=(\bs{L}_t^u, \bs{F}_t^u, \bs{A}_t^u)$. We utilize masks to distinguish between the constrained and unconstrained parts of the structure. The first $N^c$. atoms are the constrained atoms and the other atoms are unconstrained. The mask for lattice, coordinates, and atom types are $\bs{m}_{\bs{L}}$, $\bs{m}_{\bs{F}}$ and $\bs{m}_{\bs{A}}$ respectively. These masks are equal to 1 for the constrained components, which we aim to impose on the generated materials. In SCIGEN, 
$\bs{m}_{\bs{L}}$ is 1 for the $\bs{l}_{1}$ and $\bs{l}_{2}$, as these two lattice vectors define the unit cell of AL layer and they need to result in the lattice vectors we impose. On the other hand, $\bs{m}_{\bs{L}}$ is equal to 0 for $\bs{l}_{3}$, as $\bs{l}_{3}$ does not involve in the formation of AL structure directly and we can let the model to generate $\bs{l}_{3}$. For fractional coordinates, $\bs{m}_{\bs{F}}$ is equal to 1 for the atoms indexed from first to $N^c$-th atoms, which form AL layer structures. The same rule applies for $\bs{m}_{\bs{A}}$, the mask for atom types.

\newpage
\section{SCIGEN from a probability perspective}
By design, because of the masking mechanics of SCIGEN, given the constrained structure $\bs{M}_0^c$, the generated structure $\bs{M}_0$ would has $\bs{m}\odot \bs{M}_0 = \bs{m}\odot \bs{M}_0^c$. However, for the scheme to be useful, the distribution of the generated structure $P(\bs{M}_0|\bs{M}_0^c)$ needs to reflect the actual distribution of the stable material conditional on the constraints imposed. Let's assume that the trained model has $q$ as its diffusion inference model, and $p$ as its denoising generative model. For a given constrained structure $\bs{M}_0^c$, we can independently sample diffused constrained structure at every time step with
\begin{equation*}
    P(\bs{M}_t^c| \bs{M}_0^c) = q_{0,t}(\bs{M}_t^c|\bs{M}_0^c).
\end{equation*}
For compactness, let's define $\overline{\bs{M}}$ as the masked part of $\bs{M}$, $\bs{m}\odot \bs{M}$, and $\widetilde{\bs{M}}$ as the unmasked part of $\bs{M}$, $(1-\bs{m})\odot \bs{M}$. This means that 
\begin{align*}
    \bs{M}^c_t &= \overline{\bs{M}}^c_t + \widetilde{\bs{M}}^c_t \\
    \bs{M}^u_t &= \overline{\bs{M}}^u_t + \widetilde{\bs{M}}^u_t \\
    \bs{M}_t &= \overline{\bs{M}}^c_t + \widetilde{\bs{M}}^u_t
\end{align*}
Then, consider a generation of a structure $\bs{M}_0$ given $\bs{M}_t^c$ for every time step $t\in [0..T]$, $\bs{M}_{0:T}^c$, the distribution of $\bs{M}_0$ is
\begin{align*}
    P(\bs{M}_0|\bs{M}_{0:T}^c) &= P(\overline{\bs{M}}_0^c, \widetilde{\bs{M}}_0^u|\overline{\bs{M}}_{0:T}^c, \widetilde{\bs{M}}_{0:T}^c) \\
    &= P(\widetilde{\bs{M}}_0^u|\overline{\bs{M}}_{0:T}^c, \widetilde{\bs{M}}_{0:T}^c) \\
    &= \int P(\overline{\bs{M}}_0^u, \widetilde{\bs{M}}_0^u|\overline{\bs{M}}_{0:T}^c, \widetilde{\bs{M}}_{0:T}^c) d\overline{\bs{M}}_0^u \\
    &= \int p_{1,0}(\overline{\bs{M}}_0^u, \widetilde{\bs{M}}_0^u| \overline{\bs{M}}_1^c, \widetilde{\bs{M}}_1^u)\cdot P(\overline{\bs{M}}_1^c, \widetilde{\bs{M}}_1^u|\overline{\bs{M}}_{0:T}^c, \widetilde{\bs{M}}_{0:T}^c) d\overline{\bs{M}}_0^u d\widetilde{\bs{M}}_1^u \\
    &= \int p_{1,0}(\overline{\bs{M}}_0^u, \widetilde{\bs{M}}_0^u|\overline{\bs{M}}_1^c, \widetilde{\bs{M}}_1^u)\cdot P(\widetilde{\bs{M}}_1^u|\overline{\bs{M}}_{0:T}^c, \widetilde{\bs{M}}_{0:T}^c) d\overline{\bs{M}}_0^u d\widetilde{\bs{M}}_1^u \\
    &\cdots \\
    &= \int \left[\prod_{t=1}^{T}p_{t,t-1}(\overline{\bs{M}}_{t-1}^u, \widetilde{\bs{M}}_{t-1}^u|\overline{\bs{M}}_t^c, \widetilde{\bs{M}}_t^u)\right]\cdot P(\widetilde{\bs{M}}_T^u|\overline{\bs{M}}_{0:T}^c, \widetilde{\bs{M}}_{0:T}^c) d\overline{\bs{M}}_{0:T-1}^u d\widetilde{\bs{M}}_{1:T}^u \\
    &= \int \left[\prod_{t=1}^{T}p_{t,t-1}(\overline{\bs{M}}_{t-1}^u, \widetilde{\bs{M}}_{t-1}^u|\overline{\bs{M}}_t^c, \widetilde{\bs{M}}_t^u)\right]\cdot P(\widetilde{\bs{M}}_T^u) d\overline{\bs{M}}_{0:T-1}^u d\widetilde{\bs{M}}_{1:T}^u \\
    &= \int p_{1:T,0:T-1}(\overline{\bs{M}}_{0:T-1}^u, \widetilde{\bs{M}}_{0:T}^u|\overline{\bs{M}}_{1:T}^c) d\overline{\bs{M}}_{0:T-1}^u d\widetilde{\bs{M}}_{1:T}^u \\
    &= p_{1:T,0}(\widetilde{\bs{M}}_0^u| \overline{\bs{M}}_{1:T}^c).
\end{align*}
The fourth equality is from the Markov structure of the generative model. The sixth equality is from recursively apply second to fifth equalities. Since $\bs{M}_T^u$ was randomly sampled, it is independent of $\bs{M}_{0:T}^c$. Hence, the seventh equality holds. Then, the Markov structure gives the eighth equality, and marginal integration gives the last. However, this result is for a specific sequence $\bs{M}_{0:T}^c$. By taking average over $\bs{M}_{1:T}^c$ for a given constrained structure $\bs{M}_0^c$,
\begin{align*}
    P(\bs{M}_0|\bs{M}_0^c) &= \int p_{1:T,0}(\widetilde{\bs{M}}_0^u|\overline{\bs{M}}_{1:T}^c)\cdot q_{0,1:T}(\bs{M}_{1:T}^c|\bs{M}_0^c) d\bs{M}_{1:T}^c \\
    &= \int p_{1:T,0}(\widetilde{\bs{M}}_0^u| \overline{\bs{M}}_{1:T}^c)\cdot q_{0,1:T}(\overline{\bs{M}}_{1:T}^c|\overline{\bs{M}}_0^c) d\overline{\bs{M}}_{1:T}^c \\
    &\approx q_{0,0}(\widetilde{\bs{M}}_0^u| \overline{\bs{M}}_0^c)
\end{align*}
where the approximation holds if $p$ from the underlying generative model is well-trained, so that $p$ can approximate the diffusion pathway of $q$ in a reverse order. This implies that the SCIGEN structure generation given $\bs{M}_0^c$ is equivalent to conditional generation of the underlying generative model by constraining $\overline{\bs{M}}_0$ with $\overline{\bs{M}}_0^c$ and filling in $\widetilde{\bs{M}}_0$.

One might also ask whether this result would still valid when the base model contain predictor-corrector mechanism like in DiffCSP used in our demonstration since the prediction at half-time-step points require additional masking from diffused constrained structure. Similar to above, but we also given $\bs{M}_t^c$ for every half-time step $t\in [\frac{1}{2}..T-\frac{1}{2}]$, i.e., given $\bs{M}_{0:\frac{1}{2}:T}^c$,
\begin{align*}
    P(\bs{M}_0|\bs{M}_{0:\frac{1}{2}:T}^c) &= P(\bs{L}_0, \bs{F}_0, \bs{A}_0|\bs{M}_{0:\frac{1}{2}:T}^c)\\
    &= P(\bs{L}_0, \overline{\bs{F}}^c_0, \widetilde{\bs{F}}^u_0, \bs{A}_0|\bs{M}_{0:\frac{1}{2}:T}^c) \\
    &= P(\bs{L}_0, \widetilde{\bs{F}}^u_0, \bs{A}_0|\bs{M}_{0:\frac{1}{2}:T}^c) \\
    &= \int P(\bs{L}_0, \bs{F}^u_0, \bs{A}_0|\bs{M}_{0:\frac{1}{2}:T}^c) d\overline{\bs{F}}^u_0 \\
    &= \int p^{\bs{F},c}_{1,0}(\bs{F}^u_0| \bs{L}_0, \bs{F}_{\frac{1}{2}}, \bs{A}_0) P(\bs{L}_0, \bs{F}_{\frac{1}{2}}, \bs{A}_0|\bs{M}_{0:\frac{1}{2}:T}^c) d\overline{\bs{F}}^u_0 d\widetilde{\bs{F}}^u_{\frac{1}{2}} \\
    &= \int p^{\bs{F},c}_{1,0}(\bs{F}^u_0| \bs{L}_0, \bs{F}_{\frac{1}{2}}, \bs{A}_0)P(\bs{L}_0, \widetilde{\bs{F}}^u_{\frac{1}{2}}, \bs{A}_0|\bs{M}_{0:\frac{1}{2}:T}^c) d\overline{\bs{F}}^u_0 d\widetilde{\bs{F}}^u_{\frac{1}{2}} \\
    &= \int p^{\bs{F},c}_{1,0}(\bs{F}^u_0| \bs{L}_0, \bs{F}_{\frac{1}{2}}, \bs{A}_0)P(\bs{L}_0, \bs{F}^u_{\frac{1}{2}}, \bs{A}_0|\bs{M}_{0:\frac{1}{2}:T}^c) d\overline{\bs{F}}^u_0 d\overline{\bs{F}}^u_{\frac{1}{2}} d\widetilde{\bs{F}}^u_{\frac{1}{2}}\\
    &= \int p^{\bs{F},c}_{1,0}(\bs{F}^u_0| \bs{L}_0, \bs{F}_{\frac{1}{2}}, \bs{A}_0)P(\widetilde{\bs{L}}^u_0, \bs{F}^u_{\frac{1}{2}}, \widetilde{\bs{A}}^u_0|\bs{M}_{0:\frac{1}{2}:T}^c) d\overline{\bs{F}}^u_0 d\bs{F}^u_{\frac{1}{2}}\\
    &= \int p^{\bs{F},c}_{1,0}(\bs{F}^u_0| \bs{L}_0, \bs{F}_{\frac{1}{2}}, \bs{A}_0)P(\bs{L}^u_0, \bs{F}^u_{\frac{1}{2}}, \bs{A}^u_0|\bs{M}_{0:\frac{1}{2}:T}^c) d\overline{\bs{M}}^u_0 d\bs{F}^u_{\frac{1}{2}}\\
    &= \int p^{\bs{F},c}_{1,0}(\bs{F}^u_0| \bs{L}_0, \bs{F}_{\frac{1}{2}}, \bs{A}_0)p_{1,0}^{\bs{L}}(\bs{L}^u_0|\bs{M}_1)p_{1,0}^{\bs{F},p}(\bs{F}^u_{\frac{1}{2}}|\bs{M}_1)p_{1,0}^{\bs{A}}(\bs{A}^u_0|\bs{M}_1) P(\bs{M}_1|\bs{M}_{0:\frac{1}{2}:T}^c) d\overline{\bs{M}}^u_0 d\bs{F}^u_{\frac{1}{2}}d\widetilde{\bs{M}}^u_1
\end{align*}
where fifth, and last equality are from corrector, and predictor mechanisms, respectively. We can combine the product of the first four terms as $p_{1,0}(\bs{L}^u_0, \bs{F}^u_0, \bs{A}^u_0, \bs{F}^u_{\frac{1}{2}}|\overline{\bs{L}}^c_0,\overline{\bs{A}}^c_0,\overline{\bs{F}}^c_{\frac{1}{2}},\bs{M}_1) = p_{1,0}(\bs{M}^u_0, \bs{F}^u_{\frac{1}{2}}|\overline{\bs{L}}^c_0,\overline{\bs{A}}^c_0,\overline{\bs{F}}^c_{\frac{1}{2}},\bs{M}_1)$. We can apply this recursively:
\begin{align*}
    P(\bs{M}_0|\bs{M}_{0:\frac{1}{2}:T}^c) &= \int \prod_{t=0}^{T-1}\left[p_{t+1,t}(\bs{M}^u_{t}, \bs{F}^u_{t+\frac{1}{2}}|\overline{\bs{L}}^c_t,\overline{\bs{A}}^c_t,\overline{\bs{F}}^c_{t+\frac{1}{2}},\bs{M}_{t+1})\right]\cdot P(\bs{M}_T|\bs{M}_{0:\frac{1}{2}:T}^c) d\overline{\bs{M}}^u_{0:T-1} d\bs{F}^u_{\frac{1}{2}:T-\frac{1}{2}}d\widetilde{\bs{M}}^u_{1:T}\\
    &= \int \prod_{t=0}^{T-1}\left[p_{t+1,t}(\bs{M}^u_{t}, \bs{F}^u_{t+\frac{1}{2}}|\overline{\bs{L}}^c_t,\overline{\bs{A}}^c_t,\overline{\bs{F}}^c_{t+\frac{1}{2}},\bs{M}_{t+1})\right]\cdot P(\widetilde{\bs{M}}^u_T) d\overline{\bs{M}}^u_{0:T-1} d\bs{F}^u_{\frac{1}{2}:T-\frac{1}{2}}d\widetilde{\bs{M}}^u_{1:T} \\
    &= \int p_{1:T,0:T-1}(\bs{M}^u_{0:T-1}, \bs{F}^u_{\frac{1}{2}, T-\frac{1}{2}} |\overline{\bs{L}}^c_0,\overline{\bs{A}}^c_0, \overline{\bs{M}}^c_{1:T},\overline{\bs{F}}^c_{\frac{1}{2}:T-\frac{1}{2}},\widetilde{\bs{M}}^u_{T}) \cdot P(\widetilde{\bs{M}}^u_T) d\overline{\bs{M}}^u_{0:T-1} d\bs{F}^u_{\frac{1}{2}:T-\frac{1}{2}}d\widetilde{\bs{M}}^u_{1:T}\\
    &= \int p_{1:T,0:T-1}(\bs{M}^u_{0:T-1}, \bs{F}^u_{\frac{1}{2}, T-\frac{1}{2}} |\overline{\bs{L}}^c_0,\overline{\bs{A}}^c_0, \overline{\bs{M}}^c_{1:T},\overline{\bs{F}}^c_{\frac{1}{2}:T-\frac{1}{2}},\widetilde{\bs{M}}^u_{T}) \cdot P(\widetilde{\bs{M}}^u_T)d\widetilde{\bs{M}}^u_{T} d\overline{\bs{M}}^u_{0}d\bs{M}^u_{1:T-1} d\bs{F}^u_{\frac{1}{2}:T-\frac{1}{2}}\\
    &= \int p_{1:T,0:T-1}(\widetilde{\bs{M}}^u_{0}|\overline{\bs{L}}^c_0,\overline{\bs{A}}^c_0, \overline{\bs{M}}^c_{1:T},\overline{\bs{F}}^c_{\frac{1}{2}:T-\frac{1}{2}},\widetilde{\bs{M}}^u_{T}) \cdot P(\widetilde{\bs{M}}^u_T)d\widetilde{\bs{M}}^u_{T}\\
    &= p_{1:T,0}(\widetilde{\bs{M}}^u_{0}|\overline{\bs{L}}^c_0,\overline{\bs{A}}^c_0, \overline{\bs{M}}^c_{1:T},\overline{\bs{F}}^c_{\frac{1}{2}:T-\frac{1}{2}})
\end{align*}
where the second equality is from the independence of $\bs{M}^u_T$ with respected to $\bs{M}_{0:\frac{1}{2}:T}^c$. Then, the
Markov structure gives the third equality, and marginal integration gives the last two equalities. By taking average over $\bs{M}^c_{\frac{1}{2}:\frac{1}{2}:T}$ for a given constrained structure $\bs{M}^c_0$,
\begin{align*}
    P(\bs{M}_0|\bs{M}_0^c) &= \int p_{1:T,0}(\widetilde{\bs{M}}^u_{0}|\overline{\bs{L}}^c_0,\overline{\bs{A}}^c_0, \overline{\bs{M}}^c_{1:T},\overline{\bs{F}}^c_{\frac{1}{2}:T-\frac{1}{2}})\cdot q_{0,\frac{1}{2}:\frac{1}{2}:T}(\bs{M}_{\frac{1}{2}:\frac{1}{2}:T}^c|\bs{M}_0^c) d\bs{M}_{\frac{1}{2}:\frac{1}{2}:T}^c \\
    &= \int p_{1:T,0}(\widetilde{\bs{M}}^u_{0}|\overline{\bs{L}}^c_0,\overline{\bs{A}}^c_0, \overline{\bs{M}}^c_{1:T},\overline{\bs{F}}^c_{\frac{1}{2}:T-\frac{1}{2}})\cdot q_{0,\frac{1}{2}:\frac{1}{2}:T}(\overline{\bs{L}}^c_0,\overline{\bs{A}}^c_0, \overline{\bs{M}}^c_{1:T},\overline{\bs{F}}^c_{\frac{1}{2}:T-\frac{1}{2}}|\overline{\bs{M}}_0^c) d\overline{\bs{L}}^c_0 d\overline{\bs{A}}^c_0 d\overline{\bs{M}}^c_{1:T} d\overline{\bs{F}}^c_{\frac{1}{2}:T-\frac{1}{2}}\\
    &\approx q_{0,0}(\widetilde{\bs{M}}_0^u| \overline{\bs{M}}_0^c),
\end{align*}
i.e., the conclusion we have before also hold for the DiffCSP case where there is Predictor-Corrector mechanism in the denoising generative model.

\newpage

\stepcounter{suppNoteCounter}
\section{Training of the generative model}

\begin{figure}[H]
\includegraphics[width=1.0\textwidth]{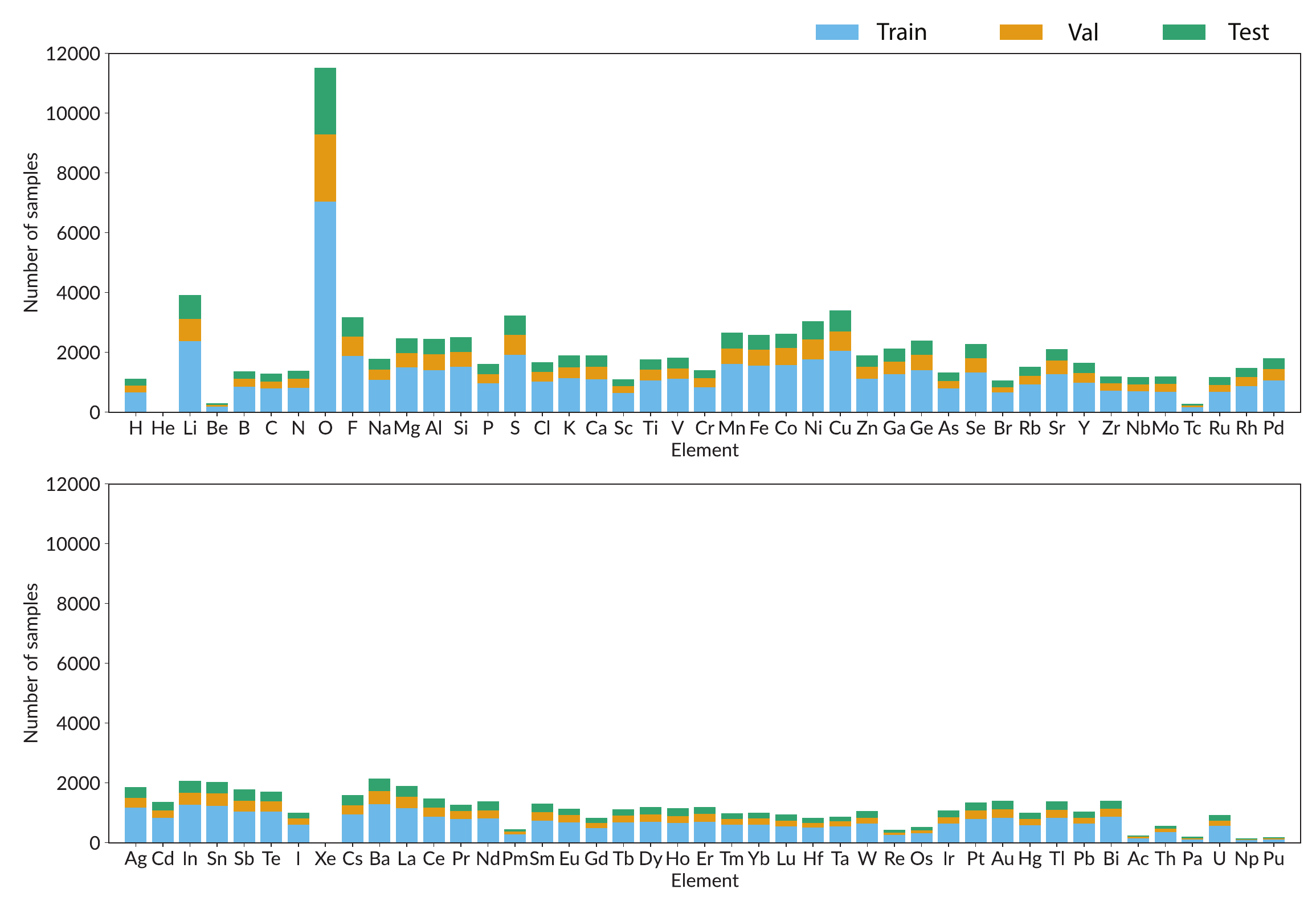}
\centering
\caption{
\textbf{Training and testing data by elements in the MP-20 dataset.}  
The number of appearances by each chemical element in the training (blue), validation (orange), and testing (green) datasets.
}
\label{SI_mp20_counts}
\end{figure}

To train DiffCSP architecture, we used the MP-20\cite{xie2018crystal, jain2013commentary} dataset of 45231 materials from the Materials Project database. MP-20 covers 89 elements, as Figure \ref{SI_mp20_counts} shows the element distribution for each of the train, validation, and test datasets. MP-20 includes most experimentally known materials with up to 20 atoms per unit cell. All materials are selected from the stable ones after DFT relaxation. MP-20 includes materials with energy above the hull ($E_{\text{hull}}$) smaller than 0.08 eV/atom and formation energy smaller than 2 eV/atom. 

\newpage

\stepcounter{suppNoteCounter}
\section{Stability pre-screening procedures of generated materials}

\begin{figure}[H]
\includegraphics[width=0.8\textwidth]{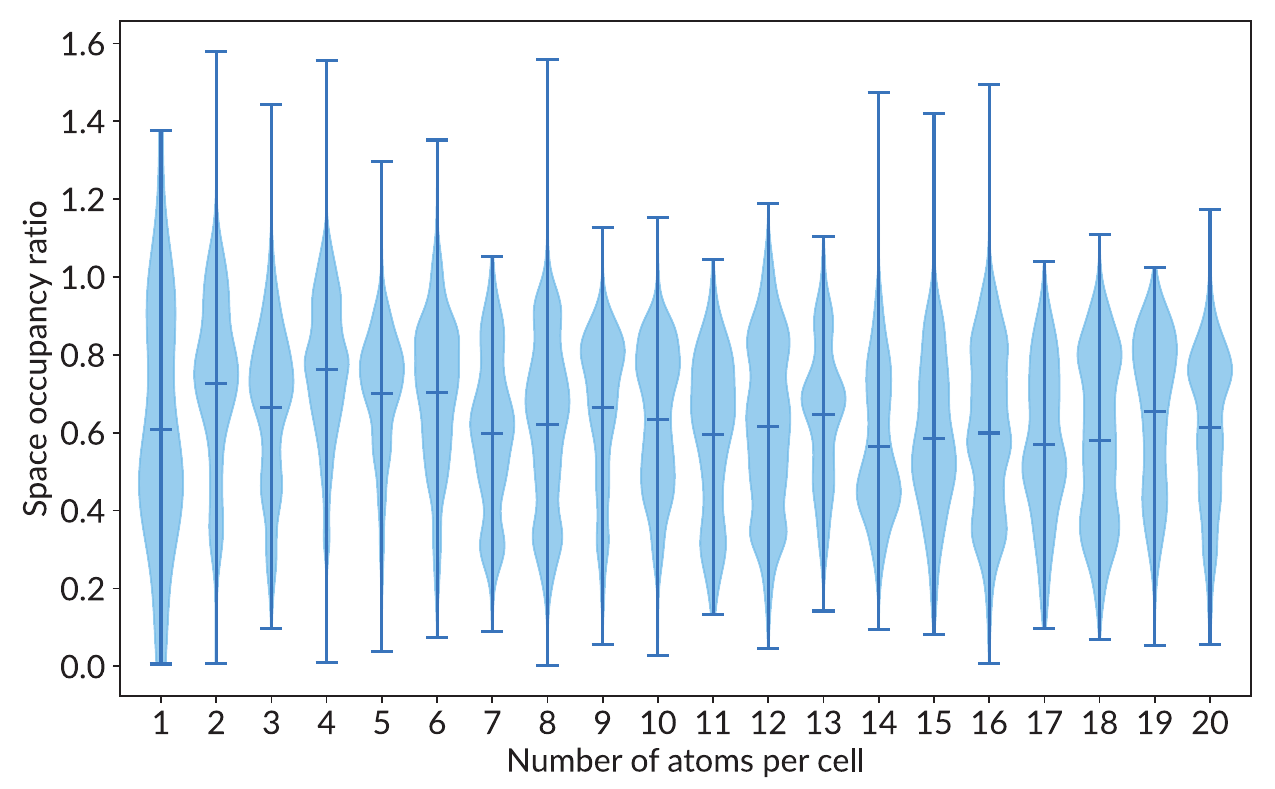}
\centering
\caption{
\textbf{Space occupancy distribution of MP-20 dataset.} 
We surveyed the space occupancy ratio of the 45229 materials in the MP-20 dataset. The highest value is 1.578.
}
\label{SI_space_occ}
\end{figure}

It is crucial to investigate whether the generated materials are stable to determine the candidate materials for density functional theory (DFT) calculations and for further practical application. In our work, we present a four-staged stability screening that is quick and efficiently removes unrealistic samples. The first two filters are straightforward: charge neutrality and the ratio of unit cell space occupied by atoms. Each crystal needs to be neutral about its electrical charge. We used the SMACT\cite{davies2019smact} library to search if each material could meet this requirement. Additionally, we considered the ratio of the space occupied by the atoms per the volume of the unit cell. As the violin plots in Fig. \ref{SI_space_occ} show, the space occupation ratio $R_{occ}$ ranges in a similar manner over different $N$ values in the MP-20 dataset. Among the materials generated by SCIGEN, some exhibit excessively high $R_{occ}$ values, indicating that too many or large atoms are confined within a narrow space of the unit cells. We set $R_0$ = 1.7 as the threshold, which is slightly higher than the maximum $R_{occ}$ for each $N$ in Figure \ref{SI_space_occ}, and discarded the materials with higher $R$ values than $R_0$.

\begin{figure}[H]
\includegraphics[width=1.0\textwidth]{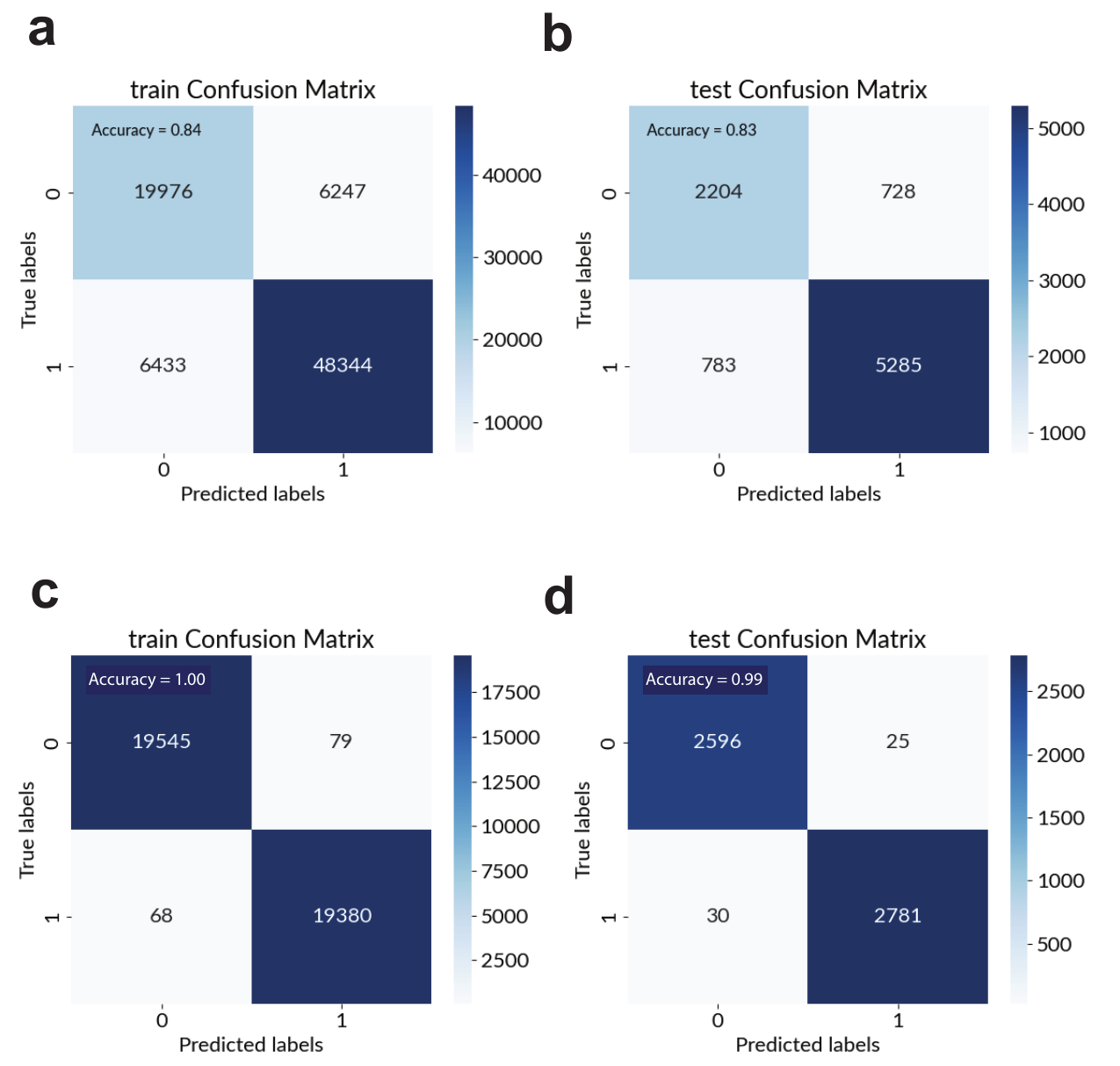}
\centering
\caption{
\textbf{Confusion matrix of GNN-based stability classifier.} 
The class labels '1' and '0' indicate stable and unstable, respectively. \textbf{a.} Training data of $E_{\text{hull}}$ classification ($\Psi_1$). \textbf{b.} Test data of $E_{\text{hull}}$ classification ($\Psi_1$). \textbf{c.} Training data of diffused structure classification ($\Psi_2$). \textbf{d.} Test data of diffused structure classification ($\Psi_2$).
}
\label{SI_gnn_cm}
\end{figure}

Next, we built classification models based on Graph Neural Networks (GNN) that take crystal structures as input. Evaluation of the crystal structure stability needs to take the atom types and their geometric relations into account, and GNNs are capable of dealing with these together. To learn the stable phases from the existing material database, we trained two GNN models, $\Psi_1$ and $\Psi_2$, as simple and fast tools. $\Psi_1$ was trained to classify whether the given materials present energy above the convex hull ($E_{\text{hull}}$) higher or lower than 0.1 eV. We set $E_{\text{hull}}$ lower than 0.1 eV as the stable samples. We accessed the materials with the $E_{\text{hull}}$ values from the Matbench Discovery\cite{riebesell2023matbench}. $\Psi_2$ was designed to evaluate if the input materials are pristine and distinctive from diffused structures. This model was trained to classify pristine materials in the MP-20\cite{xie2018crystal, jain2013commentary} dataset and those diffused by Gaussian noise with 5$\%$ deviation. We implemented GNN models $\Psi_1$ and $\Psi_2$ using E3NN library\cite{geiger2022e3nn, chen2021direct}. Figure \ref{SI_gnn_cm} shows the confusion matrices for the two GNN models, which confirm their capabilities for stability classification.

Using the criteria described above, we pre-screened the generated materials. Table \ref{tab_prescreen} provides the ratio of survived materials relative to the total generated material structures after each filter. We named the pre-screening procedures of charge neutrality, space occupation ratio, $\Psi_1$, and $\Psi_2$ as Filters 1 through 4, respectively. Approximately 6 to 8 percent of the generated materials passed the filters when we imposed the geometry of Triangular, Honeycomb, and Kagome lattice layers. Compared to DiffCSP, the survival ratio is lower by an order. This difference arises from the constraints of specific geometry within the crystal structures, which can lead to higher failure rates compared to the vanilla model without rigorous imposition of structures. However, we were able to generate new materials with AL geometries, which are rare and can potentially expand the boundaries of the existing material database.

\begin{table}[H]
\begin{center}
\caption{\textbf{Pre-screening evaluation}}
\label{tab_prescreen}
\begin{tabular}{l*{5}{c}r}
            AL & Triangular & Honeycomb & Kagome & DiffCSP \\
\hline
Filter 1 & 92.14 & 83.90 & 66.44 & 81.43 \\
Filter 2 & 86.13 & 83.15 & 65.99 & 81.43 \\
Filter 3 & 40.89 & 24.24 & 19.91 & 64.38 \\
Filter 4 & 6.13 & 7.33 & 8.35 & 62.54 \\
\hline
\end{tabular}
\end{center}
\end{table}

For the three AL types (Triangular, Honeycomb, Kagome), we tracked the ratio of materials passing the pre-screening filters at each timestamp $t \in[T, 0]$. More materials could pass the filters as the structures are denoised, or in other words, as $t$ gets smaller. This trend indicates that the denoising method optimizes the components of $\boldsymbol{M}_t=(\boldsymbol{F}_t, \boldsymbol{L}_t, \boldsymbol{A}_t)$ into suitable values. Especially with high enough $t$ close to $T$=1000, the elements of the lattice matrix $\boldsymbol{L}_t$ can be approximated to follow a normal distribution. Consequently, the volume of the unit cell is significantly smaller than that of actual crystals. Therefore, almost no samples pass the criteria of space occupation ratio (Filter 2), resulting in a narrow area below the green line with a large $t$ region.

\begin{figure}[H]
\includegraphics[width=1.0\textwidth]{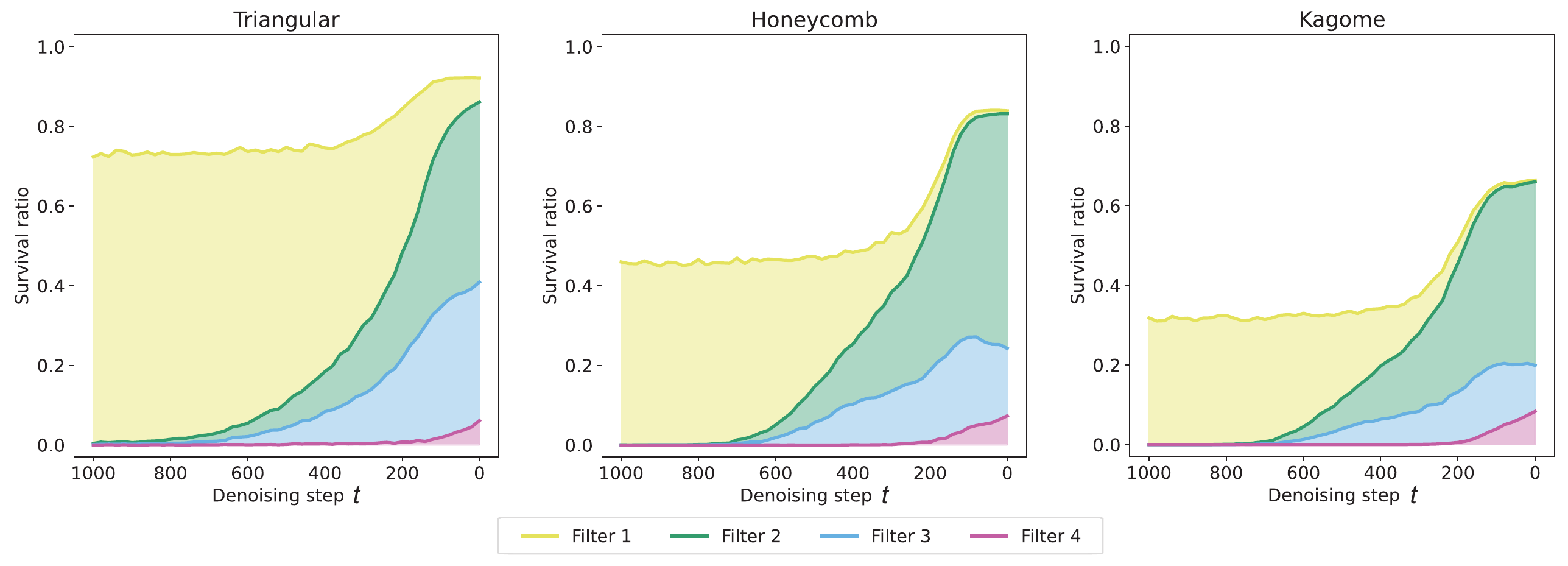}
\centering
\caption{
\textbf{The survival ratio of materials under pre-screening through denoising process.}  
Filter 1, Filter 2, Filter 3, and Filter 4 represent the criteria of charge neutrality, density, energy above the convex hull, and diffusiveness, respectively.
}
\label{SI_denoise_filter}
\end{figure}



\newpage

\stepcounter{suppNoteCounter}
\section{Generated materials with Archimedean Lattice constraints}

We present the database of the generated materials based on Archimedean lattice (AL) structures. This involves mass generation of materials, pre-screening, density functional theory (DFT) calculations, and evaluation of the stability and novelty of the generated materials.

To create the material database, we first generated materials with AL constraints using our SCIGEN framework. For each sample, we sampled the atom types as the AL vertices $\mathcal{A}^c$ from the uniform distribution of the 10 magnetic atoms (Mn, Fe, Co, Ni, Ru, Nd, Gd, Tb, Dy, and Yb). We then sampled the distances $d^c$ between the nearest neighbor vertices from the KDE in Fig. \ref{SI_bond_len}. We sampled the number of atoms per unit cell $N$ from the generated profile in Fig. \ref{SI_natm_count_total}. To regulate the complexity of generated materials, we specified $N_{\text{max}}$, the maximum value of $N$, as min(4$N^{c}$, 20) for each AL. After all, the whole dataset contains 7.87 million materials. 

We pre-screened the initially generated materials to ensure a manageable and diverse dataset for further analysis by DFT. We first applied the four-layered pre-screening filters, which we mentioned in Supplementary Information 5, to narrow down the outputs to 790 thousand. Subsequently, we employed DFT for the structure relaxation of these materials. The DFT calculations were performed on 26,000 materials down-sampled from the 790 thousand, resulting in 13.8 thousand stable phases. Stability was evaluated based on three primary metrics derived from the DFT results: maximum interatomic force after relaxation ($F_{\text{max}}$), average changes in lattice constants ($d_{\text{latt}}$), and average changes in atomic coordinates ($d_{\text{xyz}}$). For a material to be considered stable, we set the upper bound for $F_{\text{max}}$ to 0.01 eV/$\AA$. To evaluate the degree of structural changes through DFT, the thresholds for $d_{\text{latt}}$ and $d_{\text{xyz}}$ were determined specifically for each class of AL structures. To ensure the novelty of the generated materials, we filtered out duplicates from the dataset. This process involved two main steps: first, identifying structural matches with the MP-20 training dataset. We used StructureMatcher from Pymatgen\cite{ong2013python} to evaluate the correspondence of the structures. Second, we removed materials with identical structures within the generated set. When multiple outputs presented the same crystal structures, we retained the one with the lowest $d_{\text{xyz}}$ value to ensure the highest precision in atomic positioning. 

Table \ref{tab_novelty} presents the threshold values of $d_{\text{latt}}$ and $d_{\text{xyz}}$ ($\AA$) and the number of materials that remained after each filtration step. Our analysis revealed that materials with square lattice structures exhibited relatively lower novelty due to the prevalence of similar structures in the dataset, typically consisting of 2 atoms per unit cell with a cubic lattice. Conversely, other AL types demonstrated significantly higher novelty, indicating that our generative method effectively produced materials outside the known distribution. Furthermore, materials that passed all the criteria are presented in Fig. S15-S218. For Lieb-like lattice materials, we present the calculated band structures of each material structure. We cover all of the stable materials ($F_{\text{max}}$ $\textless$ 0.01 eV/$\AA$) in Table S4-S14. Note that the bold IDs in these tables mean the materials meeting the criteria of $F_{\text{max}}$, $d_{\text{latt}}$ and $d_{\text{xyz}}$, which we display the crystal structures in Fig. S15-S218. 

Finally, it is noteworthy that generating stable materials with the Great rhombitrihexagonal (\textit{grt}) lattice structures proved to be particularly challenging under the current constraints. As such, further work is required to address these difficulties and is planned for future studies.

\begin{table}[H]
\begin{center}
\caption{\textbf{The threshold values of distortion ($d_{\text{latt}}$ and $d_{\text{xyz}}$) and the counts of materials survive after DFT}}
\label{tab_novelty}
\begin{tabular}{|c||c|c||c|c|c|c|} 
\hline
Archimedean lattice & $d_{\text{latt}}$ & $d_{\text{xyz}}$ & \begin{tabular}[t]{@{}c@{}}(1) Stable\\($F_{\text{max}}$)\end{tabular} & \begin{tabular}[t]{@{}c@{}} (2) Small distortion\\($d_{\text{latt}}$, $d_{\text{xyz}}$)\end{tabular} & \begin{tabular}[t]{@{}c@{}} (3) Novel structures\\(wrt MP-20 train)\end{tabular} & (4) Remove duplicates \\
\hline
\hline
Triangular & 1.0 & 0.5 & 2584 &  1692  &  1544  &  765\\
\hline
Honeycomb & 1.0 & 0.5 & 1829 & 250  &  245  &  235  \\ 
\hline
Kagome & 1.0 & 0.5 & 1489  & 452  &  439  &  392 \\
\hline
Square & 1.0 & 0.5 & 2754 & 2363  &  1679  &  538  \\
\hline
Elongated triangular & 1.0 & 0.5 & 1356 & 127  &  127  &  125  \\
\hline
Snub square & 1.0 & 0.5 & 957 & 134  &  132  &  131  \\
\hline
Truncated square & 2.0 & 1.0 & 553 &  143  &  143  &  143  \\
\hline
Small rhombitrihexagonal & 2.0 & 1.0 & 736 & 305  &  305  &  268  \\
\hline
Snub hexagonal & 2.0 & 1.0 & 668 & 212  &  212  &  191  \\
\hline
Truncated hexagonal & 3.0 & 1.5 & 125 & 40  &  40  &  40  \\
\hline
Lieb & 1.0 & 0.5 & 829 & 213  &  205  &  190  \\

\hline
\end{tabular}
\end{center}
\end{table}

\includepdf[pages=-, pagecommand={\thispagestyle{fancy}}]{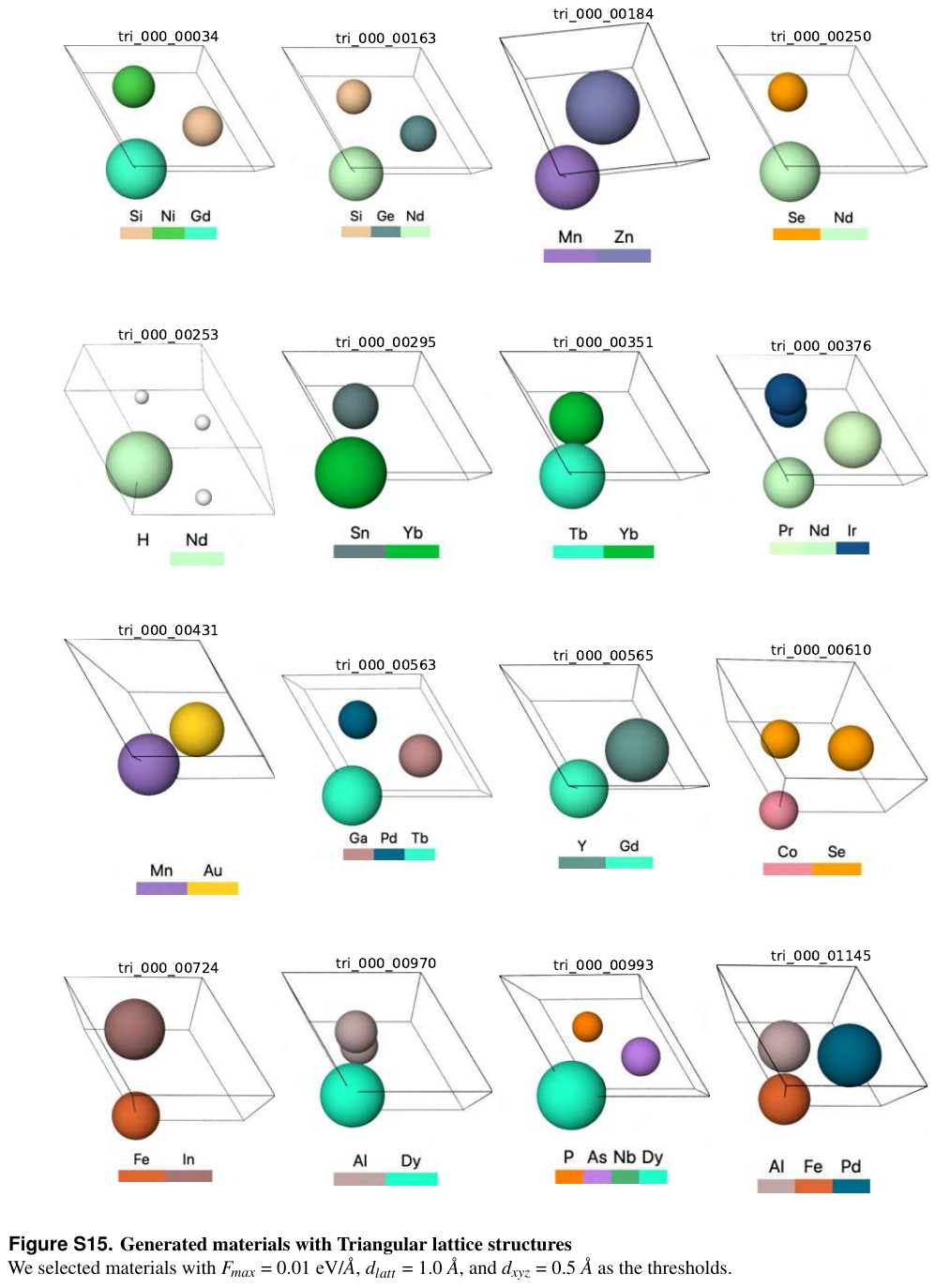}

\FloatBarrier
\printbibliography